\documentclass[twocolumn,prl,aps,twocolumn,superscriptaddress]{revtex4}
\usepackage[latin9]{inputenc}
\usepackage{amsmath}
\usepackage{amssymb}
\usepackage{graphicx}
\usepackage{makecell}
\usepackage{physics}
\usepackage{booktabs}
\usepackage[unicode=true,pdfusetitle,
 bookmarks=false,
 breaklinks=false,pdfborder={0 0 1},backref=false,colorlinks=false]
 {hyperref}
\hypersetup{
 bookmarksnumbered=false,bookmarksopen=false}

\makeatletter
\@ifundefined{textcolor}{}
{%
 \definecolor{BLACK}{gray}{0}
 \definecolor{WHITE}{gray}{1}
 \definecolor{RED}{rgb}{1,0,0}
 \definecolor{GREEN}{rgb}{0,1,0}
 \definecolor{BLUE}{rgb}{0,0,1}
 \definecolor{CYAN}{cmyk}{1,0,0,0}
 \definecolor{MAGENTA}{cmyk}{0,1,0,0}
 \definecolor{YELLOW}{cmyk}{0,0,1,0}
}

\usepackage{color}%\usepackage{amstext}

\@ifundefined{textcolor}{}{%
 \definecolor{BLACK}{gray}{0}
 \definecolor{WHITE}{gray}{1}
 \definecolor{RED}{rgb}{1,0,0}
 \definecolor{GREEN}{rgb}{0,1,0}
 \definecolor{BLUE}{rgb}{0,0,1}
 \definecolor{CYAN}{cmyk}{1,0,0,0}
 \definecolor{MAGENTA}{cmyk}{0,1,0,0}
 \definecolor{YELLOW}{cmyk}{0,0,1,0}
}

\usepackage{txfonts}

\makeatother

\begin{document}
\title{Dynamics in an exact solvable many body system: benchmark for quantum computer}

\author{Zheng-Xin Guo}
\affiliation{Guangdong-Hong Kong Joint Laboratory of Quantum Matter, Frontier Research Institute for Physics, South China Normal University, Guangzhou 510006, China}

\author{Xi-Dan Hu}
\affiliation{Guangdong Provincial Key Laboratory of Quantum Engineering and Quantum Materials, SPTE, South China Normal University, Guangzhou 510006, China}
\affiliation{GPETR Center for Quantum Precision Measurement, South China Normal University, Guangzhou 510006, China}

\author{Xue-Jia Yu}
\email{xuejiayu@pku.edu.cn}
\affiliation{International Center for Quantum Materials, School of Physics, Peking University, Beijing 100871, China}

\author{Zhi Li}
\email{lizphys@m.scnu.edu.cn}
\affiliation{Guangdong-Hong Kong Joint Laboratory of Quantum Matter, Frontier Research Institute for Physics, South China Normal University, Guangzhou 510006, China}

%\author{Shi-Liang Zhu}
%\email{slzhu@163.com}
%\affiliation{Guangdong Provincial Key Laboratory of Quantum Engineering and Quantum Materials, SPTE, South China Normal University, Guangzhou 510006, China}
%\affiliation{GPETR Center for Quantum Precision Measurement, South China Normal University, Guangzhou 510006, China}
%\affiliation{Guangdong Provincial Key Laboratory of Nuclear Science, Institute of Quantum Matter, South China Normal University, Guangzhou 510006, China}
%\affiliation{Guangdong-Hong Kong Joint Laboratory of Quantum Matter, Frontier Research Institute for Physics, South China Normal University, Guangzhou 510006, China}

\begin{abstract}
Quantum magnets are never short of novel and fascinating dynamics, yet its simulation by classical computers requires exponentially-scaled computation resources, which renders the research on large-scale many-body dynamics fiendishly difficult. In this letter, we explore the dynamic behavior of 2D large-scale ferromagnetic $J_1$-$J_2$ Heisenberg model both theoretically and experimentally. First, the analytical solution of magnon dynamics is obtained to show an obvious ballistic propagation of magnon, which is typical for quantum walk. Then, we verify the dynamic behavior of the system through numerical approach of exact diagonalization and tensor network method. We also calculate out-of-time ordered correlators and butterfly velocities among different lattice points, finding that they can well depict the competition between different couplings. Finally, a quantum walk experiment is designed and conducted on the basis of IBM programmable quantum processors, and the experimental results are in consistence with our theoretical predictions. Since the analytical results can be used, in principle, to predict the behavior of large-scale quantum many-body systems and even those infinitely large, this work will help facilitate further research on quantum walk and quantum many-body dynamics in large-scale lattice systems, guide future design of quantum computers, as well as popularize quantum computers until they are known and available to every household in the world.
\end{abstract}
\date{\today}

\maketitle

{\color{blue}\textit{Introduction.---}}Dynamical properties of the quantum many-body system are intriguing and important in condensed matter physics~\cite{AZagoskin2014,IBloch2008,LAmico2008}. Quantum walk(QW), as one of the few controllable dynamical phenomena in quantum many-body systems, promises to bring a multitude of technological applications in quantum computing and other related fields~\cite{YAharonov1993,NShenvi2003,AMChilds2009,AMChilds2013,SWZhang2003,MSRudner2009,SWZhang2011,LLehman2011,IVakulchyk2019,HChalabi2019,SChakraborty2020,LKUpreti2020,PWrzosek2020,KManouchehri2014}. In 2009, Prof. A. M. Childs et al first proved that continuous QW can be used as the computational framework of programmable quantum computers~\cite{AMChilds2009,AMChilds2013}, and the following years have witnessed the rise of QW as a hot research topic. By now, QW has been experimentally realized in a variety of platforms such as optical systems~\cite{HBPerets2008,ASchreiber2010,MABroome2010,LSansoni2012,SChakraborty2016,KKWang2019,TWu2020,YWang2020,LXiao2021p1,LXiao2021p2,KKWang2021,AGeraldi2019,TGiordani2019,SBarkhofen2018,BWang2018,CChen2018,XYXu2018,XGQiang2021,HTang2018,XGQiang2018,LXiao2017,XYXu2021}, trapped ions~\cite{PXue2009,RMatjeschk2012,HSchmitz2009,Zahringer2010,MTamura2020}, cold atomic gases~\cite{MKarski2009,CWeitenberg2011,TFukuhara2013,DXXie2020,JCarlstrom2016}, superconducting circuits~\cite{VVRamasesh2017,ZGYan2019,MGong2021,YWu2021} and so on.

On the other hand, since transport properties of quantum many-body system hinge on how fast quantum information spreads, out-of-time ordered correlators(OTOCs), which are commonly used to describe thermalization and information scrambling, have aroused extensive attention in recent yearse~\cite{MCheneau2012,JZhang2017,MCTran2020,CMonroe2021,PIlzhofer2021}. As a quantum version of the classical Poisson Bracket, OTOCs can accurately reflect the growth rate of operators' noncommutativity over time. Recently, OTOCs have been used to depict information propagation in more and more systems, inclusive of quantum many-body dynamic systems~\cite{JLi2017,MGarttner2017,ANahum2018,VKhemani2018,SVSyzranov2018,MHeyl2018,MGarttner2018,BYoshida2019,CMurthy2019,KALandsman2019,SLXu2020,SChoi2020,RJLewis-Swan2020}, localization systems~\cite{KXWei2018,SLXu2019,MMcGinley2019,PRoushan2017,HTShen2017,YHuang2016,KSlagle2017,MCTran2019,KXu2018,NYYao2014}, quantum chaotic systems~\cite{EBRozenbaum2017,IGarcia-Mata2018,BBertini2019,BYan2020,DEParker2019,JRammensee2018,TRXu2020,WLZhao2021} and black hole systems~\cite{DARoberts2015,JMaldacena2016,JMaldacena1999,JdeBoer2018}. Note that, Prof. J. Maldacena, S. H. Shenker and D. Stanford theoretically proved, for the first time, the existence of propagation limit (exponential upper boundary) in black hole system and quantum Sachdev-Ye-Kitaev (SYK) model~\cite{JMaldacena2016}. This study also verified the duality between Anti-de Sitter spacetime and conformal field theory~\cite{JMaldacena1999,JdeBoer2018}. Meanwhile, the measurement of OTOCs has been realized in multiple table-top platforms~\cite{JLi2017,MGarttner2017,ANahum2018,VKhemani2018,SLXu2020,SChoi2020,RJLewis-Swan2020,CBDag2019a,CBDag2019b,KXWei2018,MCTran2019}. Today, OTOC is among the most desirable observables in describing the properties of information propagation in dynamic systems.

Though a lot of research has been done on the stationary state problem of many-body systems, dynamics of large-scale lattice systems remains a territory unsufficiently charted. This letter will be devoted to the study of dynamic properties in large-scale quantum magnets.

%%%%%%
%%%%%%
\begin{figure*}[tbhp] \centering
\includegraphics[width=0.96\textwidth]{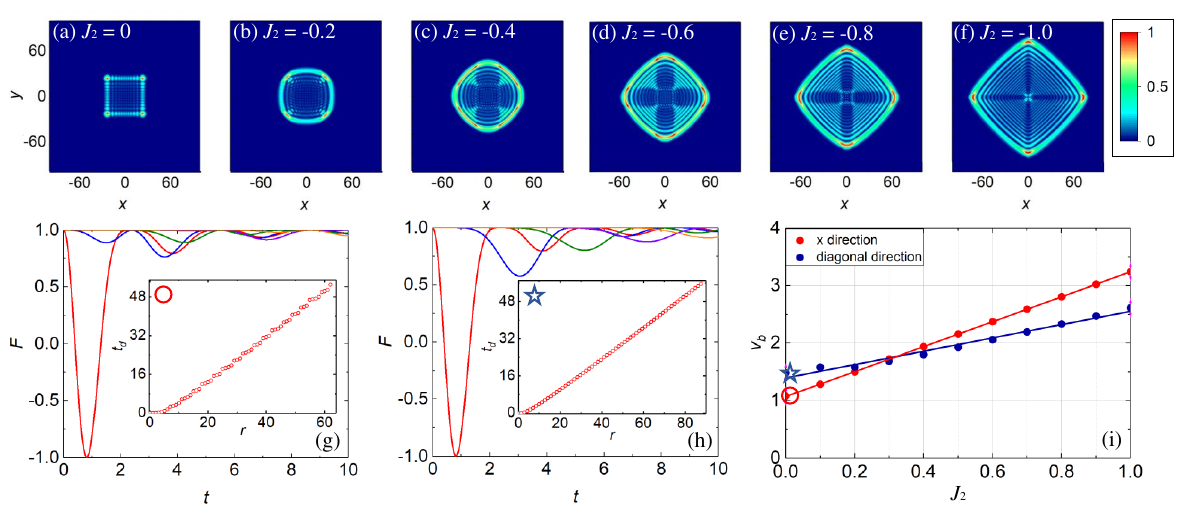}
\caption{(Color online). Visualized magnon QW and spread properties of ferromagnetic $J_1$-$J_2$ Heisenberg model with nn and nnn interaction. (a-f) The density distributions of magnon with different nnn interaction strengths $J_2$ at $t=25$. (g) The OTOC along $x$ direction for different distance $|\mathbf{r}|=\mathbf{r}_{m,0}-\mathbf{r}_0=0$ (red), $2$(blue), 4(green), 6(purple), $8$(orange). (h) The OTOC along diagonal direction for different $|\mathbf{r}|=\mathbf{r}_{m,n=m}-\mathbf{r}_0=0$ (red), $2\sqrt{2}$(blue), $4\sqrt{2}$(green), $6\sqrt{2}$(purple), $8\sqrt{2}$(orange). Insets: The $t_d$ lines (line of first decline moments) for $x$-axis and diagonal. (i) Butterfly velocities for different nnn interaction. Throughout, $J_1=-1$. The unit of $x$ and $y$ is the lattice constant a=1, whereas the time unit here is $\hbar/|J_1|=1$ with $\hbar=1$.}\label{Fig1dyn}
\end{figure*}
%%%%%%
%%%%%%

{\color{blue}\textit{Visualized QW: A. Exact Solvable Model ---}}The model adopted in this letter is a square-lattice ferromagnetic $J_1$-$J_2$ Heisenberg model, and the related Hamiltonian can be written as
\begin{equation}\label{J1J2}
\begin{split}
\hat{H}=&J_{1}\sum_{i,j}(\hat{\mathbf{S}}_{i,j}\hat{\mathbf{S}}_{i\pm1,j}+\hat{\mathbf{S}}_{i,j}\hat{\mathbf{S}}_{i,j\pm1})\\ &+J_{2}\sum_{i,j}\hat{\mathbf{S}}_{i,j}\hat{\mathbf{S}}_{i\pm1,j\pm1},
\end{split}
\end{equation}
where $J_1$ and $J_2$ correspond to the nearest-neighboring (nn) and the next-nearest-neighboring (nnn) spin exchange coupling, respectively. For simplicity, we consider only the ferromagnetic case, i.e., $J_{1}<0$ and $J_{2}<0$.

As a spin-1/2 model, it is strictly equivalent to the hard-core Bose Hubbard model which restricts the number of bosons on each site to only zero or one (see supplementary materials~\cite{suppM} for detailed process of mathematical proof). Thus, we can exactly obtain equivalent Hamiltonian as
% Besides, it is also a ferromagnetic system with the single spin-flip on top of the ground state fully equivalent to a free boson in the vacuum state. Then, by performing Fourier transformation for Eq.~(\ref{J1J2}), Hamiltonian of the hard-core boson (magnon) in diagonal form can be obtained as
%%%
%%%
\begin{equation}\label{HFM}
\begin{split}
 \hat{\widetilde{H}}=&\sum_{\mathbf{k}}\omega_{\mathbf{k}}\hat{a}^{\dagger}_{\mathbf{k}}\hat{a}_{\mathbf{k}}+J_1\sum_{i,j}(\hat{n}_{i,j}\hat{n}_{i\pm1,j}+\hat{n}_{i,j}\hat{n}_{i,j\pm1})\\&+J_2\sum_{i,j}(\hat{n}_{i,j}\hat{n}_{i\pm1,j\pm1}),
\end{split}
\end{equation}
%%%
%%%
where $\omega_{\mathbf{k}}$ stands for the energy of non-interacting magnons, which has the following expressions,
%%%
%%%
\begin{equation}\label{omega1k}
\omega_{\mathbf{k}}=2[J_{1}(1-\gamma_{1\mathbf{k}})+J_{2}(1-\gamma_{2\mathbf{k}})],
\end{equation}
%%%
%%%
%%%
%%%
\begin{equation}\label{gamma1}
  \gamma_{1\mathbf{k}}=\frac{1}{2}(\cos{k_x}+\cos{k_y}),
\end{equation}
%%%
%%%
%%%
%%%
\begin{equation}\label{gamma2}
  \gamma_{2\mathbf{k}}=\frac{1}{2}[\cos{(k_{x}+k_{y})}+\cos{(k_{x}-k_{y})}].
\end{equation}
$\hat{n}_{i,j}$ is the particle number operator and the other two terms of the Hamiltonian denote the two-body interaction between nn and nnn sites.\\

%%%
%%%

{\color{blue}\textit{B. Single Spin-Flip Dynamics and Magnon's QW---}}Then, we would like to study the propagation of a single spin-flip on top of the FM ground state. Let's start with calculating the density distribution versus time, which is a key observable easy to measure experimentally in studying QW, with its expression as~\cite{TFukuhara2013,CWeitenberg2011,MKarski2009},
%%%
%%%
\begin{equation}\label{definerou}
\begin{split}
\xi(\mathbf{r},t)=&\langle\phi|\hat{S}^{+}_{\mathbf{r}_{0}}\hat{S}^{z}_{\mathbf{r}_{m,n}}(t)\hat{S}^{-}_{\mathbf{r}_{0}}|\phi\rangle,\\
\end{split}
\end{equation}
%%%
%%%
with
%%%
%%%
\begin{equation}\label{definerou2}
\begin{split}
\hat{S}^{z}_{\mathbf{r}_{m,n}}(t)=&e^{iHt/\hbar}\hat{S}^{z}_{\mathbf{r}_{m,n}}e^{-iHt/\hbar},
\end{split}
\end{equation}
%%%
%%%
where $|\phi \rangle$ is the initial state, i.e., the ferromagnetic ground state and $\mathbf{r}=\mathbf{r}_{m,n}-\mathbf{r}_{0}$. In the expression, $\hat{S}^{z}_{\mathbf{r}_{m,n}}$ represents the spin operator at $\mathbf{r}_{m,n}$ in $z$-direction, while $\hat{S}^+_{\mathbf{r}_{0}}$ and $\hat{S}^-_{\mathbf{r}_{0}}$ denote the spin flip up and down operators at $\mathbf{r}_{0}$, respectively.

As FM system, the single spin-flip on top of the fully polarized ground state is totally equivalent to a single free hard-core boson in the vacuum state. Thus, we can simply neglect the interacting terms of Hamiltonian and apply an effective one as
\begin{equation}\label{J1J2}
\hat{H}_{eff}=\sum_{\mathbf{k}}\omega_{\mathbf{k}}\hat{a}^{\dagger}_{\mathbf{k}}\hat{a}_{\mathbf{k}}.
\end{equation}

By simple calculations, we obtain the analytical expression of magnon's evolution (see supplementary material~\cite{suppM} for details), i.e.,
%%%
\begin{equation}\label{}
\xi(\mathbf{r},t)=\left|\sum_{\mathbf{k}}\frac{e^{-i(\mathbf{k}\cdot\mathbf{r}+\omega_{\mathbf{k}}t/\hbar)}}{N^2}\right|^{2}.
\end{equation}
%%%

% As FM system, the propagation of free magnon in the vacuum state can be viewed as the spread of single spin-flip in system's ground state.
For convenient, we call the above technique as analytical single spin-flip dynamics (ASSD) from now on. The process of magnon's dynamic evolution under different parameters $J_2$ is visualized in Fig.~\ref{Fig1dyn}(a-f). As shown in the figure, when $J_2=0$, the system degenerates to the standard Heisenberg model. The density distribution then features ballistic propagation, typical for QW. When $J_2\neq0$, the nnn coupling comes into play. As shown in Fig.~\ref{Fig1dyn}, the influence of the nnn coupling on dynamics of the system can be distilled into three points:

I. $J_2$ causes the magnon to travel faster in all directions. As we know, the computation speed of quantum computers is closely related to the speed of information propagation between qubits. In other words, the acceleration of information propagation is an effective way to speed up quantum computation.

II. The acceleration is anisotropic, maximum in the $x$- and $y$-axis direction and minimum in the diagonal direction. Therefore, new quantum computers can be designed on the basis of the anisotropic propagation, which will play a pivotal role in conducting anisotropic computation and quantum simulation effectively.

III. The competition between nn and nnn coupling cannot destroy the ballistic propagation behavior. However, the geometric structure of ballistic propagation can be changed by manipulating parameters $J_1$ and $J_2$. As shown in Fig.~\ref{Fig1dyn}(a-f), the system transforms from a square structure to a rhombic QW.\\

{\color{blue}\textit{C. OTOCs and Butterfly Velocities---}}Furthermore, we calculate OTOCs of the system and the corresponding butterfly velocity to reveal the propagation characteristics~\cite{JLi2017,MGarttner2017,ANahum2018,VKhemani2018,SVSyzranov2018,MHeyl2018,MGarttner2018,BYoshida2019,CMurthy2019,KALandsman2019,SLXu2020,SChoi2020,RJLewis-Swan2020}. Note that, butterfly velocity and the Lieb-Robinson bounds (the upper limit of the information propagation speed in quantum systems) are highly interdependent, which reflects information propagation behavior in many-body systems~\cite{EHLieb1972,SBravyi2006,BNachtergaele2006,MCheneau2012,NLashkari2013,JJunemann2013,MFoss-Feig2015,DMStorch2015,CCChien2015,TMatsuta2017,DVElse2020,TKuwahara2020}. Detailed derivation of OTOCs and butterfly velocity in $J_1$-$J_2$ model is included in Supplementary Material~\cite{suppM}.

By simple operator gymnastics (ASSD), one can obtain the analytic expression of the system's OTOCs, which reads
%%%
\begin{equation}
\hat{F}(t)=\langle \phi|\hat{S}^{+}_{\mathbf{r}_0}\hat{S}_{\mathbf{r}_{m,n}}^z(t)\hat{S}_{\mathbf{r}_0}^z(0)\hat{S}_{\mathbf{r}_{m,n}}^z(t)\hat{S}_{\mathbf{r}_0}^z(0)\hat{S}^{-}_{\mathbf{r}_0} |\phi\rangle.
\end{equation}
%%%
Then, we get OTOCs' analytical expression as
%%%
\begin{equation}
\hat{F}(t)=1-8/N^2\Omega_1\Omega_2+8/N^4\Omega_1\Omega_2\Omega_1\Omega_2,
\end{equation}
%%%
where,
%%%
\begin{equation}
\begin{split}
\Omega_1=&\sum_{\mathbf{k}}{e^{-i\mathbf{k}\cdot(\mathbf{r}_0-\mathbf{r}_{m,n})}}e^{i\omega_{\mathbf{k}}t/\hbar},\\
\Omega_2=&\sum_{\mathbf{k}}{e^{i\mathbf{k}\cdot(\mathbf{r}_0-\mathbf{r}_{m,n})}}e^{-i\omega_{\mathbf{k}}t/\hbar}.
\end{split}
\end{equation}
%%%
Based on the ASSD techniques, the propagation properties of magnetic information (a spin-flip signal) in the large-scale $J_1$-$J_2$ Heisenberg model can be revealed at a lower computational cost.

The results of OTOCs and the butterfly velocity are plotted in Fig.~\ref{Fig1dyn}(g-i). We define the time when OTOC begins to decline as the critical time, denoted as $t_{d}$. The illustration reflects the relationship between $t_d$ and $\mathbf{r}$. As shown in Fig.~\ref{Fig1dyn}(g) [(h)], the OTOCs along the $x,y$-axis [diagonal] direction oscillate versus time, where the amplitude of OTOCs decreases and $t_d$ increases with the increasing $\mathbf{r}$. Since the magnon wave packet itself is conservative in the process of evolution, the probability of magnon distribution at a distant point is relatively small, hence the relatively small amplitude~\cite{MCheneau2012,JZhang2017,MCTran2020,suppM}. Besides, the propagation mode of the system is determined by the growth of the local Heisenberg operator, therefore, the longer the propagation time of the distant lattice magnon, the larger the corresponding $t_d$. The above two points well reflect that the magnon information diffuses in a way of QW. Note that, the slope of $t_d$ in the illustration is the butterfly velocity~\cite{EBRozenbaum2017,IGarcia-Mata2018,BBertini2019}. Further, we calculate how the butterfly velocity of the system changes with $J_2$ [see Fig.~\ref{Fig1dyn}(i)]. As shown in the figure, when $J_2$ increases, the butterfly velocity in the diagonal direction will slowly surpass that in the $x$-direction, which indicates that the butterfly velocity can very well reflect the competitive relationship between nn coupling and nnn coupling in the system.

In short, based on OTOCs and butterfly velocity, one can understand the features of magnon QW in $J_1$-$J_2$ model from the following three perspectives:

I. The superposition of another QW on top of the previous one will lead to the acceleration of information propagation;

II. The competition between the two results in the change from diagonal dominance to x- and y-axis dominance of QW, and correspondingly, the geometric structure changes from square to rhombus;

III. The introduction of nnn coupling will not destroy the ballistic propagation mode, therefore, the system still displays QW.

It is noticeable that the QW and OTOCs results of a single spin-flip is totally accurate, which is supported by mathematical proofs in the previous sections and double-checked by different numerical methods in supplementary material~\cite{suppM}. \\

{\color{blue}\textit{Digital quantum simulation---}}In principle, ASSD theory can very effectively describe the diffusion process of information in large-scale or even infinite lattice ferromagnetic system, thus makes it an ideal way to test the computational power of quantum computers. Although at present the several largest superconducting circuits quantum computers in the world total less than 70 quantum bits~\cite{MGong2021,YWu2021}, we believe that the coming decades will see a great increase in the number of qubits, as well as the realization of error-correcting qubits. With the unstoppable progress of technological infrastructure, more phenomena in many-body dynamics, such as QW in large-scale systems and controllable magnetic transport, will be verified by more programmable quantum processors in the future.

As a principle testing, here we conduct experiments on a 5-qubit quantum processor and compare the experimental results with the theoretical ones. In order to conduct an experiment of scrambling dynamics on the IBM quantum processor, one must translate the time evolution into the language of quantum circuits, i.e., a series of standard quantum gates. First, for single spin-flip dynamics, the effective Heisenberg model described in bosonic language can be written as
%%%
\begin{equation}\label{bosonicH}
H=-\frac{1}{2}N+2\sum_{l}(\hat{a}^{\dagger}_{l}\hat{a}_{l})-\sum_{l}(\hat{a}_{l}\hat{a}^{\dagger}_{l+1}+\hat{a}^{\dagger}_{l}\hat{a}_{l+1}).
\end{equation}
%%%
Here we take $J_1=-1$ and $S=1/2$. By ignoring the constant term, which only brings in a global phase factor, the time evolution operator can be written as
%%%
\begin{equation}\label{Time_evol}
  \hat{U}(\delta t)=e^{-iH\delta t}=e^{-i[2\sum_{l}(\hat{a}^{\dagger}_{l}\hat{a}_{l})-\sum_{l}(\hat{a}_{l}\hat{a}^{\dagger}_{l+1}+\hat{a}^{\dagger}_{l}\hat{a}_{l+1})]\delta t}.
\end{equation}
%%%
For tiny $\delta t$, the Trotter decomposition can be used to break the evolutionary operator into the following steps,
%%%
\begin{equation}\label{Time_evol_trot}
\begin{split}
  \hat{U}(\delta t) & \simeq e^{-i\sum_{l}(\hat{a}^{\dagger}_{l}\hat{a}_{l})\delta t} e^{i\sum_{l}(\hat{a}_{l}\hat{a}^{\dagger}_{l+1}+\hat{a}^{\dagger}_{l}\hat{a}_{l+1})\delta t} e^{-i\sum_{l}(\hat{a}^{\dagger}_{l}\hat{a}_{l})\delta t} \\
    & =\hat{U}_{1}(\delta t)\hat{U}_{2}(\delta t)\hat{U}_{1}(\delta t).
\end{split}
\end{equation}
%%%
%%%
where $\hat{U}_1$ and $\hat{U}_2$ represent the contribution of the on-site term and nn term. Next, we need to implement the $\hat{U}_1$ and $\hat{U}_2$ operators with a series of single and double qubit logic gates. $\hat{U}_1$ can be readily implemented via a single qubit logic gate, while $\hat{U}_2$ can be formed by the combination of C-NOT gate, Y gate and Z gate, as shown in Fig.~\ref{fig2ex}(a). The input parameter $\theta$ is a function of $\delta t$ and $J_1$. In the experiment, $\delta t=0.1$ and $J_1=-1$. Thus, a complete life cycle of time-evolving operation can be written as the quantum circuit in Fig.~\ref{fig2ex}(b).

%%%%%%
%%%%%%
\begin{figure}[tbhp] \centering
	\includegraphics[width=0.48\textwidth]{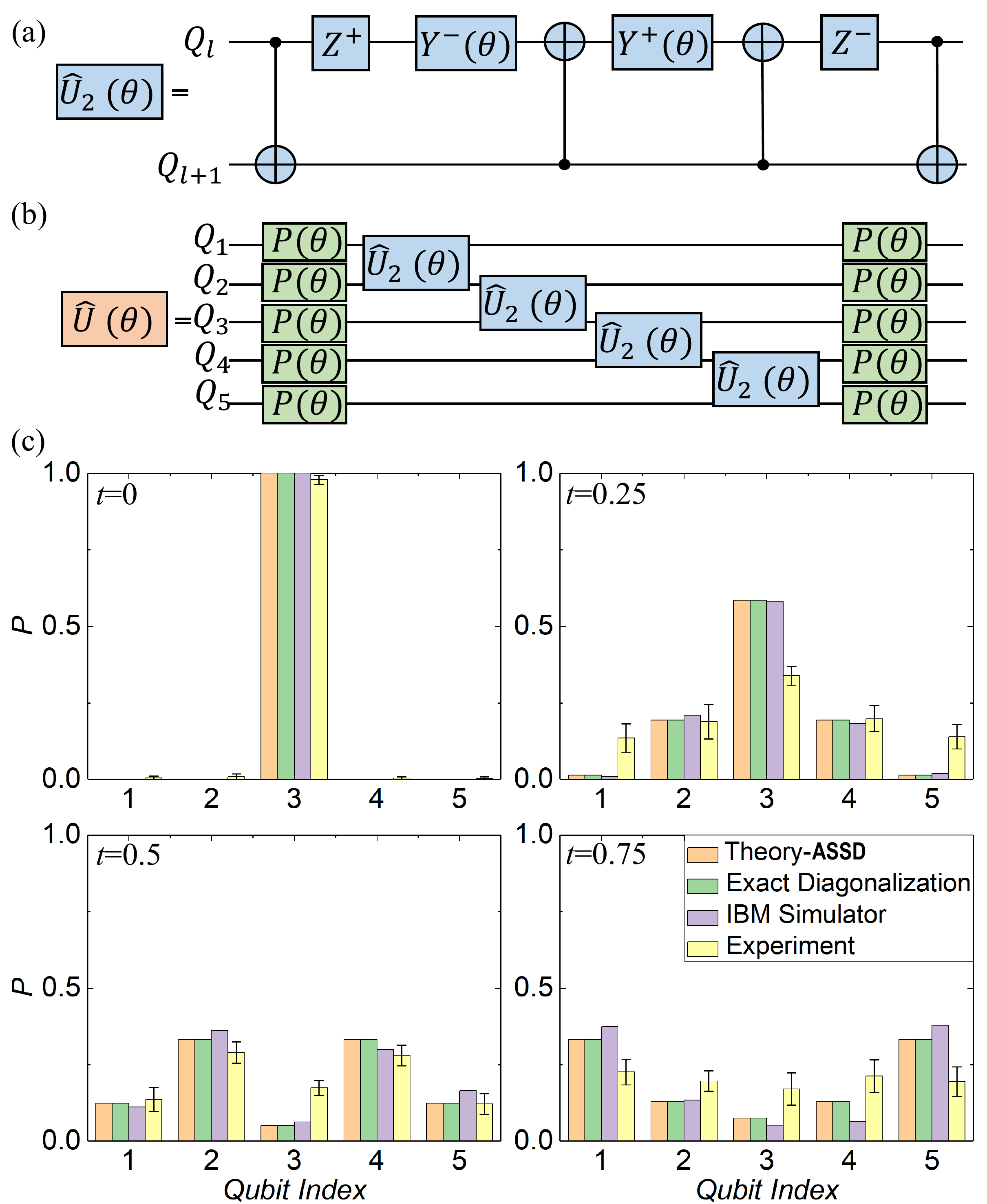}
	\caption{(Color online). The illustration of quantum circuit. (a) The quantum circuit of two-qubit time-evolution operation, where $Z^{\pm}=e^{\mp i(3\pi/2)\sigma_{z}}$ and $Y^{\pm}=e^{\mp i(\theta/2)\sigma_{y}}=e^{\mp i(\delta t/2)\sigma_{y}}$. (b) The quantum circuit which realizes the complete time-evolution operation, where $P(\theta)$ is the standard phase transition gate and $\theta=\delta t$. (c) The single spin-flip dynamics of theoretical and experimental approaches when $t=0$, $t=0.25$, $t=0.5$, and $t=0.75$, respectively.}\label{fig2ex}
\end{figure}
%%%%%%
%%%%%%

Fig.~\ref{fig2ex}(c) shows the magnon dynamic evolution behavior obtained via four different methods (analytical solution of ASSD theory, exact diagonalization method, IBM simulator and real IBM quantum processor). The results obtained by all these methods are in consistence, i.e., the system exhibits the feature of QW. It is worth noting that the analytical solution calculated by SW stands out as the ideal choice, for it can predict the dynamic evolution behavior of large-scale magnetic models at a lower computational cost of classical computers.

Since the analytical solution given here has universality, it can be used as a ``standard" to measure the performance and fidelity of large-scale quantum chips in the future. Currently, since quantum computers are limited in fidelity and number of qubits, only small-scale models can be simulated. Recent experiments with quantum circuits, however, have shown more and more controllable qubits~\cite{MGong2021,YWu2021}. In the future, with the increase of the number of qubits and the advent of error-correcting qubits, more phenomena of many-body dynamics can be studied using the method presented in this paper.\\

{\color{blue}\textit{Conclusion.---}}The propagation speed of the information is a key problem in many-body dynamics, quantum computation and black hole evolution. In this letter, we study the characteristics of quasiparticle propagation and the information scrambling process in the system by using the $J_1$-$J_2$ Heisenberg model as an example. The results show that the competition between nn and nnn will not destroy the defining features of QW, however, it does change how magnon diffuses in the system so much as to accelerate the propagation of information. In addition, we show that OTOCs and the butterfly velocity can be well used to depict the competition between nn and nnn within the system and the corresponding propagation behavior. Finally, we realize digital quantum simulation of QW on IBM quantum processors.

To facilitate the repetition of our experimental results, we choose the smallest IBM 5-qubit processor. Besides, a detailed logic gate design and specific experimental process are given in the text and supplementary materials. Therefore, readers of this letter, with or without formal training in physics, can learn how to use programmable quantum computers. We hope that by doing so, quantum computer, which still remains an unattainable mirage to the general public, will gradually step down from the altar and find its way into millions of households.\\

{\color{blue}\textit{Note added.--}}After the completion of this work, we notice that Google has recently conducted a relevant research on OTOCs and information scrambling via 53-qubits~\cite{XMi2021}. Besides, IBM has increased the number of programmable qubits to 127~\cite{IBM127news}. The rapid progress in supporting technology will help verify the theory in this letter and facilitate further exploration in large-scale quantum magnets.\\

{\color{blue}\textit{Acknowledgements.--}}This work was supported by the NSFC (Grants No. 11704132), the NSAF (Grant No. U1830111), the Natural Science Foundation of Guangdong Province (No. 2021A1515012350), the KPST of Guangzhou (Grant No. 201804020055). G.Z.X. and X.D.H. contributed equally to this work.

\pagebreak
\clearpage
%\newpage
%\clearpage
%\widetext
\begin{center}
\textbf{\large Supplemental Material}
\end{center}
%%%%%%%%%% Merge with supplemental materials %%%%%%%%%%
%%%%%%%%%% Prefix a "S" to all equations, figures, tables and reset the counter %%%%%%%%%%
%%%%%%%%%% Merge with supplemental materials %%%%%%%%%%

\maketitle
%\pagebreak
%\clearpage
%\newpage
%\clearpage
%\widetext

%%%%%%%%%% Merge with supplemental materials %%%%%%%%%%
%%%%%%%%%% Prefix a "S" to all equations, figures, tables and reset the counter %%%%%%%%%%
%%%%%%%%%% Merge with supplemental materials %%%%%%%%%%
%%%%%%%%%% Prefix a "S" to all equations, figures, tables and reset the counter %%%%%%%%%%
\setcounter{equation}{0} \setcounter{figure}{0} \setcounter{table}{0}
\setcounter{page}{1} \makeatletter \global\long\def\theequation{S\arabic{equation}}
 \global\long\def\thefigure{S\arabic{figure}}
 \global\long\def\bibnumfmt#1{[S#1]}
 \global\long\def\citenumfont#1{S#1}
%\tableofcontents{}

\section*{Contents}
{\color{blue}\textit{S1. Details of exact solvable model}}

{\color{blue}\textit{S2. Details of magnon dynamics}}

{\color{blue}\textit{S3. Out-of-time ordered correlators  and butterfly velocity}}

{\color{blue}\textit{S4. The mechanics of quantum walk}}

{\color{blue}\textit{S5. Comparison of ASSD method with numerical exact diagonalization}}

{\color{blue}\textit{S6. Comparison of ASSD method with numerical Tensor Network (TDVP)}}

{\color{blue}\textit{S7. Details about the digital quantum simulation}}

{\color{blue}\textit{S8. Visualized acceleration of quantum walk}}

\section*{S1. Details of exact solvable model}
Standard $J_1$-$J_2$ Heisenberg model Hamiltonian reads
%%%
%%%
\begin{equation}
\begin{split}
\hat{H}=&J_{1}\sum_{m,n}(\hat{\mathbf{S}}_{m,n}\hat{\mathbf{S}}_{m\pm1,n}+\hat{\mathbf{S}}_{m,n}\hat{\mathbf{S}}_{m,n\pm1})\\ &+J_{2}\sum_{m,n}\hat{\mathbf{S}}_{m,n}\hat{\mathbf{S}}_{m\pm1,n\pm1},
\end{split}
\end{equation}
%%%
%%%
As it is a spin-half system, there are only two permitted states on each site, i.e., $\ket{\uparrow}$ and $\ket{\downarrow}$. Thus, the effects of applying spin operators to the states are
\begin{equation}
    \hat{S}^+ \ket{\downarrow}=\ket{\uparrow},  \hat{S}^+ \ket{\uparrow}=0,
\end{equation}
\begin{equation}
    \hat{S}^- \ket{\uparrow}=\ket{\downarrow},  \hat{S}^- \ket{\downarrow}=0,
\end{equation}
\begin{equation}
    \hat{S}^z \ket{\downarrow}=-\frac{1}{2} \ket{\downarrow}, \hat{S}^z \ket{\uparrow}=\frac{1}{2} \ket{\uparrow}.
\end{equation}
Hard-core Boson model describes a special category of bosonic system, which only allows one or zero boson on each sites, represented by $\ket{1}$ and $\ket{0}$. Similarly, the operating effects of creation and annihilation operators are
\begin{equation}
    \hat{a} \ket{1}=\ket{0}, \hat{a} \ket{0}=0,
\end{equation}
\begin{equation}
    \hat{a}^{\dagger} \ket{0}=\ket{1}, \hat{a}^{\dagger} \ket{1}=0,
\end{equation}
\begin{equation}
\hat{n} \ket{1}=\hat{a}^{\dagger}\hat{a}\ket{1}=\ket{1},\hat{n} \ket{0}=\hat{a}^{\dagger}\hat{a}\ket{0}=0.
\end{equation}
Considering the similar operation effects on states, it is obvious that the Heisenberg Hamiltonian can be equivalent to a hard-core Bosonic Hamiltonian via the transformation
\begin{equation}\label{HP}
\begin{split}
  &\hat{S}_{m,n}^{+}=\hat{a}_{m,n},\\
  &\hat{S}_{m,n}^{-}=\hat{a}^{\dagger}_{m,n},\\
  &\hat{S}_{m,n}^{z}=1/2-\hat{a}^{\dagger}_{m,n}\hat{a}_{m,n},
\end{split}
\end{equation}
%%%
%%%
Then one can obtain the Hamiltonian as,
%%%
%%%
\begin{equation}\label{HFMFULL}
\begin{split}
 \hat{\widetilde{H}}=&\sum_{\mathbf{k}}\omega_{\mathbf{k}}\hat{a}^{\dagger}_{\mathbf{k}}\hat{a}_{\mathbf{k}}+J_1\sum_{i,j}(\hat{n}_{i,j}\hat{n}_{i\pm1,j}+\hat{n}_{i,j}\hat{n}_{i,j\pm1})\\&+J_2\sum_{i,j}(\hat{n}_{i,j}\hat{n}_{i\pm1,j\pm1}),
\end{split}
\end{equation}
%%%
%%%
where $\omega_{\mathbf{k}}$ stands for the energy of non-interacting magnons and we have
%%%
%%%
\begin{equation}\label{omega1k}
\omega_{\mathbf{k}}=2[J_{1}(1-\gamma_{1\mathbf{k}})+J_{2}(1-\gamma_{2\mathbf{k}})],
\end{equation}
%%%
%%%
%%%
%%%
\begin{equation}\label{gamma1}
  \gamma_{1\mathbf{k}}=\frac{1}{2}(cosk_x+cosk_y),
\end{equation}
%%%
%%%
%%%
%%%
\begin{equation}\label{gamma2}
  \gamma_{2\mathbf{k}}=\frac{1}{2}(cos(k_{x}+k_{y})+cos(k_{x}-k_{y})).
\end{equation}
$\hat{n}_{i,j}$ is the particle number operator and the second(third) term represent two-body interaction between nn(nnn) sites.
%%%
%%%

\section*{S2. Details of magnon dynamics}
The time-dependent density $\xi$ is defined as
%%%
%%%
\begin{equation}\label{definerou}
\begin{split}
\xi(\mathbf{r},t)=&\langle\phi|\hat{S}^{+}_{\mathbf{r}_{0}}\hat{S}^{z}_{\mathbf{r}_{m,n}}(t)\hat{S}^{-}_{\mathbf{r}_{0}}|\phi\rangle,\\
\hat{S}^{z}_{\mathbf{r}_{m,n}}(t)=&e^{iHt/\hbar}\hat{S}^{z}_{\mathbf{r}_{m,n}}e^{-iHt/\hbar},
\end{split}
\end{equation}
%%%
%%%
Using the same Paradigm of the previous section, we can rewrite the contraction in the hard-core bosonic language, i.e.,
%%%
%%%
\begin{equation}\label{rouHP}
\begin{split}
  \xi(\mathbf{r},t)&=\langle\phi|\hat{a}_{\mathbf{r}_{0}}\hat{n}_{\mathbf{r}_{m,n}}(t)\hat{a}^{\dagger}_{\mathbf{r}_{0}}|\phi\rangle,\\
    \hat{n}_{\mathbf{r}_{m,n}}(t)& =e^{iHt/\hbar}\hat{n}_{\mathbf{r}_{m,n}}e^{-iHt/\hbar},
\end{split}
\end{equation}
%%%
%%%
where $\hat{n}_{\mathbf{r}_{m,n}}=\hat{a}^{\dagger}_{\mathbf{r}_{m,n}}\hat{a}_{\mathbf{r}_{m,n}}$ is the particle number operator. We can then naturally rewrite the expression of the density distribution as
%%%
%%%
\begin{equation}\label{rewriteRou}
    \xi(\mathbf{r},t)=\langle\Psi(\mathbf{r},t)|\Psi(\mathbf{r},t)\rangle,
\end{equation}
%%%
%%%
with
%%%
%%%
\begin{equation}
  |\Psi(\mathbf{r},t)\rangle=\hat{a}_{\mathbf{r}_{m,n}}e^{-iHt/\hbar}\hat{a}^{\dagger}_{\mathbf{r}_{0}}|\phi\rangle.
\end{equation}
%%%
%%%
By using Fourier transformation, we have
%%%
%%%
\begin{equation}\label{psift}
  |\Psi(\mathbf{r},t)\rangle=\frac{1}{N^2}\sum_{\mathbf{k,k}'}e^{-i(\mathbf{kr}_{m,n}-\mathbf{k}'\mathbf{r}_{0})}\hat{a}_{\mathbf{k}}e^{-iHt/\hbar}\hat{a}^{\dagger}_{\mathbf{k}'}|\phi\rangle.
\end{equation}
%%%
%%%
Since this is a FM system, the single spin-flip on top of the fully polarized FM ground state can be exactly mapped to a single hard-core boson in the vacuum state. As it known to all, in FM case, the Hamiltonian ignoring the interacting terms is still exact for single-magnon sectors. Hence, we can substitute the effective Hamiltonian
\begin{equation}\label{J1J2}
\hat{H}_{eff}=\sum_{\mathbf{k}}\omega_{\mathbf{k}}\hat{a}^{\dagger}_{\mathbf{k}}\hat{a}_{\mathbf{k}}
\end{equation}
into Eq.(\ref{psift}). Considering the operation rules of creation and annihilation operators and substituting the diagonal hamiltonian, one can obtain the analytical expression of wavefunction, i.e., %(Note that, $|\phi_{a}\rangle$ is the ground state, so we will get zero if we apply annihilation operator on it.)
%%%
%%%
\begin{equation}
\begin{split}
  |&\Psi(\mathbf{r},t)\rangle=\frac{1}{N^2}\sum_{\mathbf{k}} e^{-i(\mathbf{kr}+\omega_{\mathbf{k}}t/\hbar)}|\phi\rangle,
\end{split}
\end{equation}
where $\mathbf{r}=\mathbf{r}_{m,n}-\mathbf{r}_{0}$. By substituting it into Eq.(\ref{rewriteRou}), the density distribution $\xi(\mathbf{r},t)$ can be obtained. The analytical expression reads
%(We neglect the constant term, since it provides none qualitative changes to the dynamical properties.) Then one
%%%
%%%
\begin{equation}\label{}
\xi(\mathbf{r},t)=\left|\sum_{\mathbf{k}}\frac{e^{-i(\mathbf{kr}+\omega_{\mathbf{k}}t/\hbar)}}{N^2}\right|^{2}.
\end{equation}
%%%
%%%

\section*{S3. Out-of-time ordered correlators  and butterfly velocity}
The general definition of four-point correlation function out-of-time ordered correlators(OTOCs) is
%%%
%%%
\begin{equation}
\hat{F}(t)=\langle \hat{W}^{\dagger}(t)\hat{V}^{\dagger}(0)\hat{W}(t)\hat{V}(0) \rangle,
\end{equation}
%%%
%%%
where,
%%%
%%%
\begin{equation}
\hat{W}(t)=e^{i\hat{H}t/\hbar}\hat{W}(0)e^{-i\hat{H}t/\hbar}.
\end{equation}
%%%
%%%
In the initial state, $\hat{W}$ and $\hat{V}$ are commutative since they are operators of different positions. With time evolution, however, $\hat{W}$ can be written as the form of Baker-Campbell-Hausdorff formula,
%%%
%%%
\begin{equation}
\hat{W}(t)=\sum_{k=0}^{\infty}\frac{(it)^k}{k!}[\hat{H},...,[\hat{H},\hat{W}],...],
\end{equation}
%%%
%%%
where the high order terms in the above expression lose commutability as time goes, and gradually dominate OTOCs in the evolution process. So OTOCs show a decline versus time, i.e.,
%%%
%%%
\begin{equation}
F(t)=c_1-c_2e^{\lambda_L(t-|\mathbf{r}|/v_b)},
\end{equation}
%%%
%%%
Here, $v_b$ is the so-called butterfly velocity which describes the growth rate of a local operator versus time. The butterfly velocity in a chaotic system can define the chaotic boundary closely related to the Lieb-Robinson bound, i.e., light cone of the chaotic system. Here, we choose $\hat{W}=\hat{S}_{\mathbf{r}_{m,n}}^{z}$ and $\hat{V}=\hat{S}_{\mathbf{r}_{0}}^{z}$, where $|\mathbf{r}|$ denotes the distance between two sites, while $c_1$ and $c_2$ are constants. Then, the OTOCs become
%%%
%%%
\begin{equation}
\hat{F}(t)=\langle \phi|\hat{S}^{+}_{\mathbf{r}_0}\hat{S}_{\mathbf{r}_{m,n}}^z(t)\hat{S}_{\mathbf{r}_0}^z(0)\hat{S}_{\mathbf{r}_{m,n}}^z(t)\hat{S}_{\mathbf{r}_0}^z(0)\hat{S}^{-}_{\mathbf{r}_0} |\phi\rangle.
\end{equation}
%%%
%%%
which can help effectively understand the information spread behavior in quantum systems.

%
%The $J_1$-$J_2$ QW can be explained in a more intuitive way. For an FM system only with nn coupling, the real space Hamiltonian of the system can be described by bosonic language as
%%%%
%\begin{equation}\label{}
%\hat{H}=\sum_{m,n,m',n'}P_{mnm'n'}\hat{a}^{\dagger}_{m,n}\hat{a}_{m',n'},
%\end{equation}
%%%%
%where the probability of spin-exchange reads
%%%%
%\begin{equation}\label{pij}
%\begin{split}
%P_{mnm'n'}=&J_{1}\left(2\delta_{m=m',n=n'}+\frac{1}{2}\delta_{m=m'\pm1,n=n'}+\frac{1}{2}\delta_{m=m',n=n'\pm1}\right)\\
%&+J_{2}\left(2\delta_{m=m',n=n'}+\frac{1}{2}\delta_{m=m'\pm1,n=n'\pm1}\right).\\
%\end{split}
%\end{equation}
%%%%
%Fixing the nn term, we plot the magnons' probability distribution $P_{mnm'n'}$ under different nnn interaction in Fig.2. As for $J_2=0$, the spin-exchange coupling can only occurs between the four nearest sites [see Fig.~\ref{fig2Pmn}(a)], which is in consistence with the standard QW in Heisenberg model. This is why the QW exhibit a square structure and the transport front appear at the square's corners after a period of evolution. Furthermore, the introduction of nnn coupling will increase the probability in the diagonal directions, finally resulting in a hollow square $P_{mnm'n'}$ structure [see Fig.~\ref{fig2Pmn}(b-f)]. That leads to the rhombus QW.

\section*{S4. The mechanics of quantum walk}
One can understand the quantum walk(QW) and information spread process of the $J_1$-$J_2$ Heisenberg system from the following three perspectives.

First, QW behavior of the system can be explained based on the results of OTOCs and butterfly velocity. The propagation velocity of a single spin-flip information in the $J_1$-$J_2$ model will be determined by the competition of direction-dependent spin exchange coupling. When $|J_2|$ is increasing, the corresponding propagation velocity in the axial direction accelerates much faster compared with that in the diagonal direction, leading to the change of the information propagation velocity.

\begin{figure}[tbhp] \centering
	\includegraphics[width=0.45\textwidth]{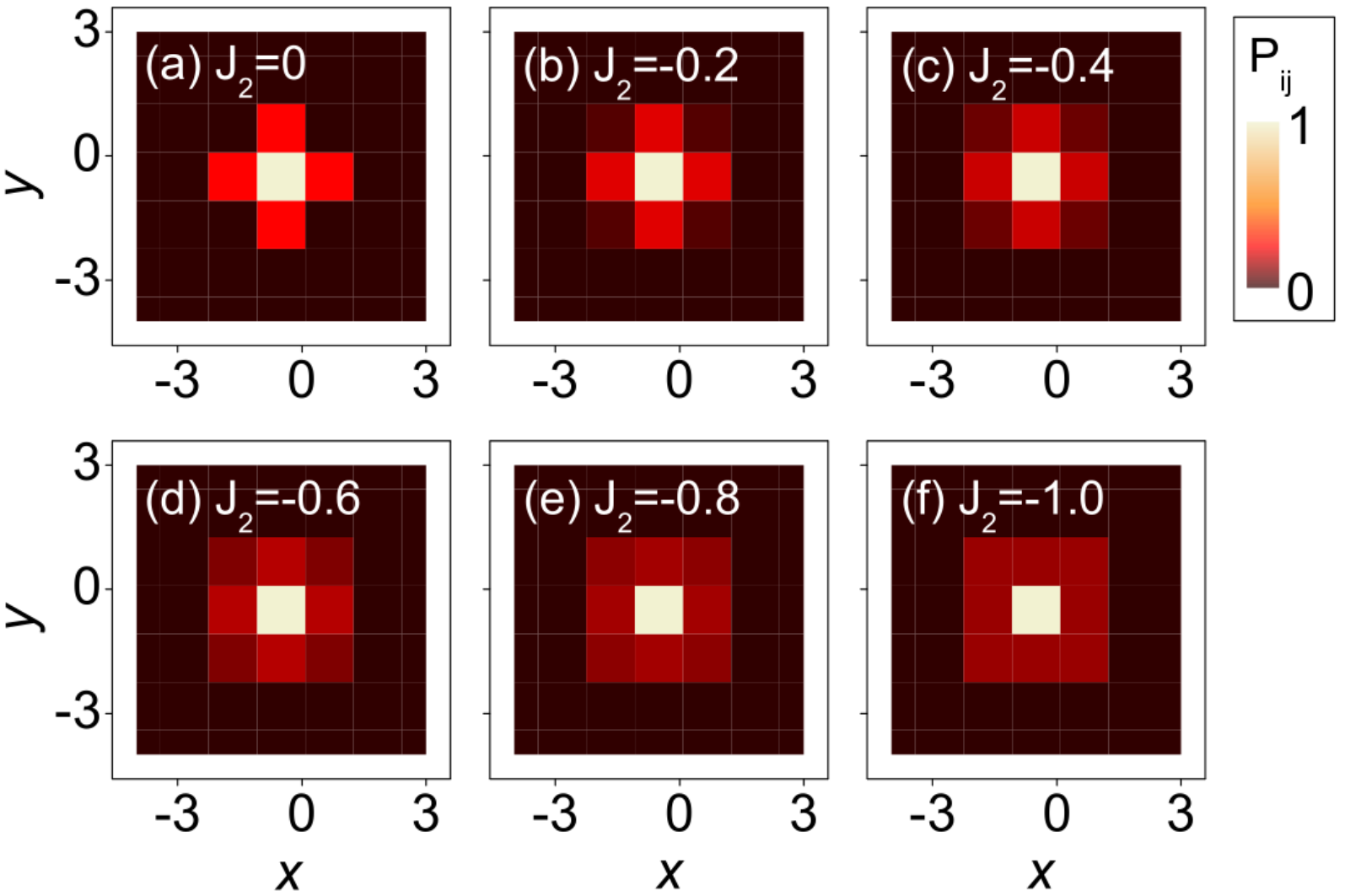}
	\caption{(Color online). The spin-exchange probability $P_{mnm'n'}$ of a single magnon in FM system with different $J_2$. The corresponding parameters are marked. }
	\label{figS1Pmn}
\end{figure}

Second, the $J_1$-$J_2$ QW can be explained in a more intuitive way. For an FM system only with nearest-neighboring(nn) coupling, the real space Hamiltonian of the system can be described by bosonic language as
%%%
\begin{equation}\label{}
\hat{H}=\sum_{m,n,m',n'}P_{mnm'n'}\hat{a}^{\dagger}_{m,n}\hat{a}_{m',n'},
\end{equation}
%%%
where the probability of spin-exchange reads
%%%
\begin{equation}\label{pij}
\begin{split}
P_{mnm'n'}=&J_{1}\left(2\delta_{m=m',n=n'}+\frac{1}{2}\delta_{m=m'\pm1,n=n'}+\frac{1}{2}\delta_{m=m',n=n'\pm1}\right)\\
&+J_{2}\left(2\delta_{m=m',n=n'}+\frac{1}{2}\delta_{m=m'\pm1,n=n'\pm1}\right).\\
\end{split}
\end{equation}
%%%
Fixing the nn term, we plot the magnons' probability distribution $P_{mnm'n'}$ under different next-nearest-neighboring(nnn) interaction in Fig.~\ref{figS1Pmn}. As for $J_2=0$, the spin-exchange coupling can only occur between the four nearest sites [see Fig.~\ref{figS1Pmn}(a)], which is in consistence with the standard QW in Heisenberg model. This is why the QW exhibits a square structure and the transport front appears at the square's corners after a period of evolution. Furthermore, the introduction of nnn coupling will increase the probability in the diagonal directions, finally resulting in a hollow square $P_{mnm'n'}$ structure [see Fig.~\ref{figS1Pmn}(b-f)]. That leads to the rhombus QW.
%%%%%%
%%%%%%
%%%%%%
%%%%%%

\begin{figure*}[tbhp] \centering
	\includegraphics[width=0.9\textwidth]{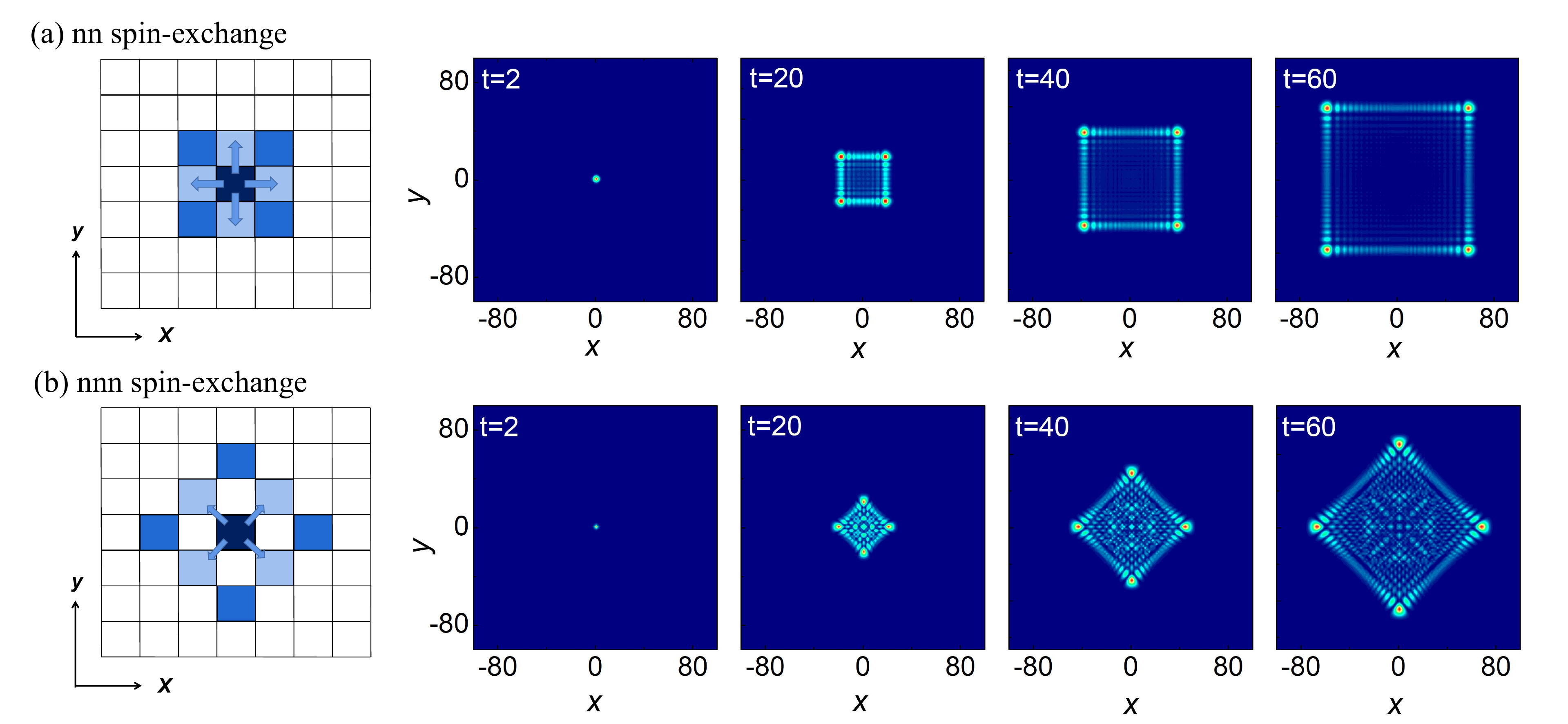}
	\caption{(Color online). The propagations properties and time-dependent density distribution of a magnon in FM case with only nn (a, $J_1=-1$ and $J_2=0$) and nnn (b, $J_1=0$ and $J_2=-1$) spin-exchange.}\label{figS2spin-ex}
\end{figure*}

Third, we can also provide a mathematical structural explanation for the QW phenomenon. When we only consider the spin-exchange coupling between nn sites($J_2=0$), the expression of $\omega_{1\mathbf{k}}$ will degenerate into $\omega_{1\mathbf{k}}=J_{1}[2-\cos{(kx)}-\cos{(ky)}]$, resulting in a square QW structure. The nnn term, however, equals to adding an operation of $J_{2}[2-\cos{(kx+ky)-\cos{(kx-ky)}}]$ on the previous basis. To be specific, it can be regarded as an operation of $45^{\circ}$ rotation plus a spacing stretching on the lattice. While it also keeps the QW behavior, the propagation structure here changes from a square to rhombus. For clearness, we plot Fig.~\ref{figS2spin-ex} to show the evolution characteristics versus time considering only the contribution of nn [Fig.~\ref{figS2spin-ex}(a)] and nnn [Fig.~\ref{figS2spin-ex}(b)] interaction. The fact that the highest density is distributed at the four corners denotes that it is actually a form of ballistic diffusion, with the highest density distributed at the spread front, instead of the classical wave mode (denser inside rather than outside).
%%%%%%
%%%%%%
%%%%%%
%%%%%%

In a word, one can easily understand the QW dynamics by using OTOCs and butterfly velocities, i.e.,

I. Another type of QW (induced by nnn term) will be superimposed on the previous one (induced by nn term), leading to an nontrivial acceleration.

II. The competition between two QW leads to structure changing from square to a rhombus.

III. The ballistic propagation (QW mechanism) keeps.

\section*{S5. Comparison of ASSD method with numerical exact diagonalization}
%%%%%%
%%%%%%
\begin{figure*}[tbhp]  \centering
	\includegraphics[width=0.9\textwidth]{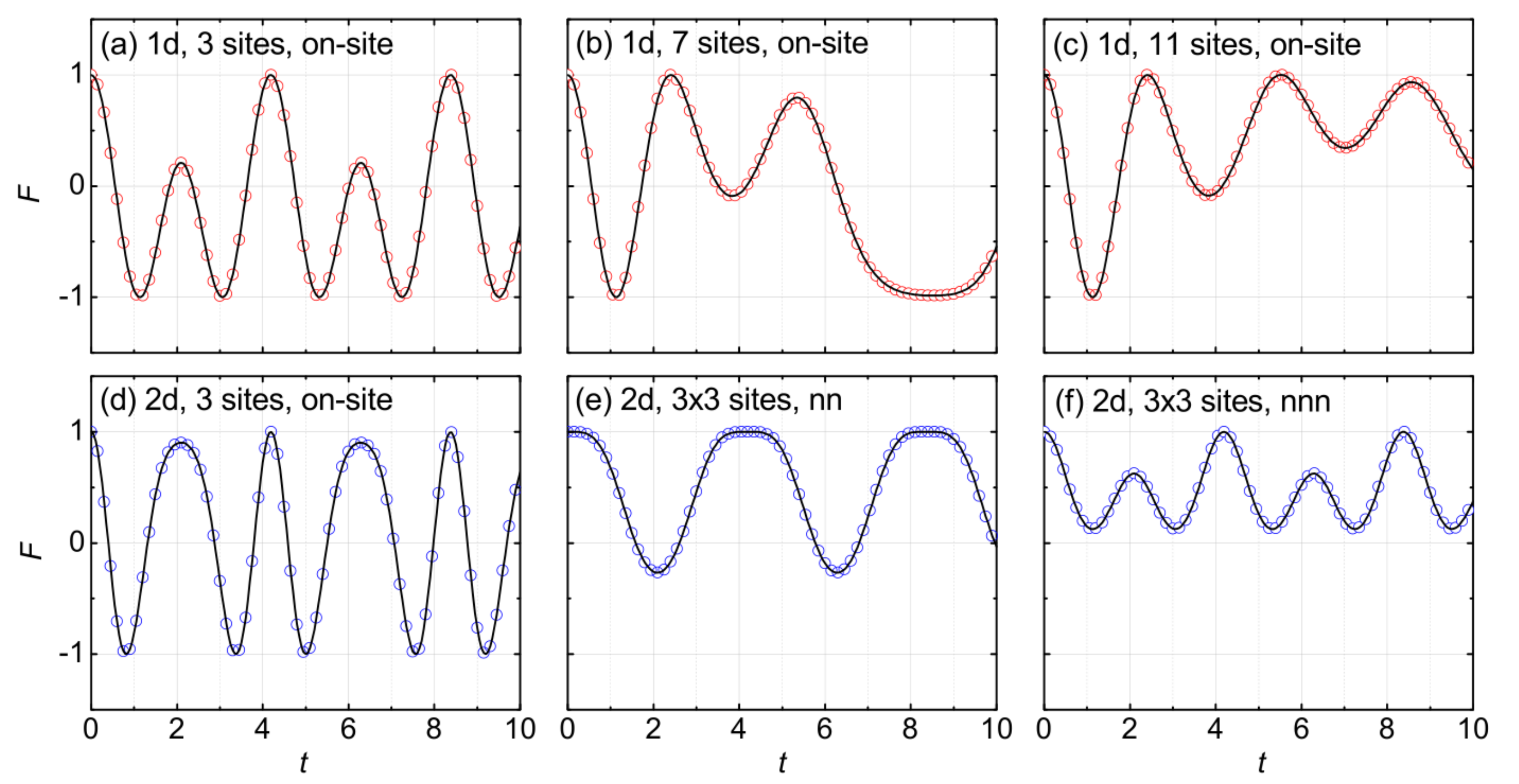}
	\caption{ (Color online). The comparison between the results of ASSD and ED (lines: ASSD results, dots: ED results). (a)$\sim$(c): Comparison of OTOC result when $\mathbf{r}_{m,n}=\mathbf{r}_0$ in one-dimensional system with 3, 7 and 11 sites, respectively. (d)$\sim$(f): Comparison of OTOC result in two-dimensional systems with 9 sites, when $\mathbf{r}_{m,n}$ is the on-site, the nn and the nnn site of the spin-flipped position.}\label{figS3otoc}
\end{figure*}
%%%%%%
%%%%%%
In this section, we verify the correctness and accuracy of ASSD theory. Size, dimension and changes in $\hat{W}$ and $\hat{V}$ operators are discussed respectively in how they are affecting OTOCs of the system. As verification, we compare the results with the exact diagonalization. First, we prove the accuracy of ASSD method in describing information spread of FM system. We compared the above analytical results with the results of exact diagonalization(ED), which are shown in Fig.~\ref{figS3otoc}.

Next, we turn to the two-dimensional case. Fig.~\ref{figS3otoc}(d-f) corresponds to $\mathbf{r}_{m,n}=\mathbf{r}_0$ (on-site), $\mathbf{r}_0+x$ (nn) and $\mathbf{r}_0+x+y$ (nnn), respectively. By comparing with the exact diagonalization results, we prove that SW theory is highly effective and accurate not only in describing high dimensional system, but also in depicting OTOCs of the coordinate axis and the diagonal direction.

\section*{S6. Comparison of ASSD method with numerical Tensor Network (TDVP)}
\begin{figure*}
    	\includegraphics[width=0.8\textwidth]{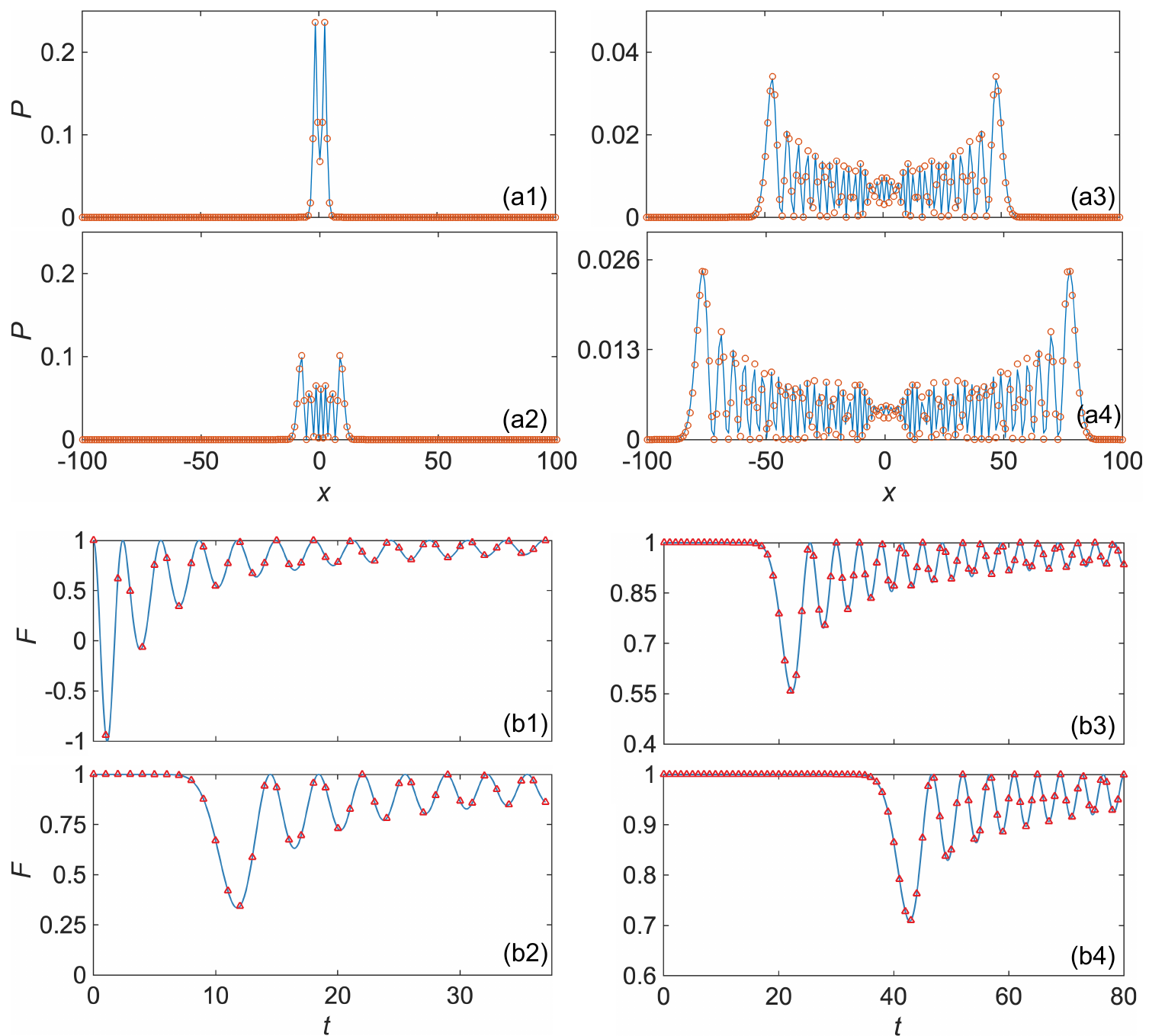}
	\caption{(Color online). The comparison of QW and OTOCs in 200-spin chain by ASSP and DQVP . (a1)-(a4): The QW results in 200-spin chain by ASSD(blue line) and TDVP(red circle) when evolution time $t=3, 10, 50$ and $80$, respectively. (b1)-(b4): The OTOCs results in 200-spin chain by ASSD(blue line) and TDVP(red triangle) when the distance between the flipped spin and the detected spin $r=r_i-r_0=0, 10, 20$ and $40$, respectively.}\label{TDVP}
\end{figure*}

In this section, we go on verifying the effectiveness of ASSD in a larger one dimension system with the help of one tensor network method----TDVP, which can provide approximate solutions in system with hundreds of sites after calculating for several days. We compare the results of QW and OTOCs from ASSD and TDVP (see ~\ref{TDVP}) and find that TDVP results are in good agreement with that of TSSP in larger systems despite of some numerical errors induced by the rise of entanglement when calculation time increases. This test further prove the efficiency and accuracy of ASSD.

\section*{S7. Details about the digital quantum simulation}
A. Quantum circuit for spin-flip dynamics\\

B. Observing quantum walk on IBM cloud quantum computer\\

C. Full data of the measured density distribution\\

\section*{A. Quantum circuit for spin-flip dynamics}
As shown in the main text, the quantum circuit for realizing digital quantum simulation(DQS) of 1D spin-flip propagation is constituted by the time evolution quantum gate $\hat{U}(\delta t)$, which is decided by the system hamiltonian
\begin{equation}
\hat{H}=-J\sum_{l} \hat{\bm{S}}_{l}\cdot\hat{\bm{S}}_{l+1}.
\end{equation}
It can also be represented in bosonic language through the standard Holstein-Primakoff transformation, which reads
\begin{equation}
\hat{H}=-0.5N+2\sum_{l} \hat{a}_{l}^{\dagger}\hat{a}_{l}-\sum_{l} (\hat{a}_l\hat{a}_{l+1}^{\dagger}+\hat{a}_{l}^{\dagger}\hat{a}_l).
\end{equation}
Then, we can obtain the explicit form of time evolution operator
\begin{equation}
\begin{split}
\hat{U}(\delta t)&=e^{-i\hat{H}\delta t}\\
&=e^{-i[2\sum_{l} \hat{a}_{l}^{\dagger}\hat{a}_{l}-\sum_{l} (\hat{a}_l\hat{a}_{l+1}^{\dagger}+\hat{a}_{l}^{\dagger}\hat{a}_l)]\delta t}
\end{split}
\end{equation}
where, $\delta t$ is a tiny period and we neglect the global phase caused by the constant term, which will not influence the probability density distribution. To demonstrate the effect of this operation via quantum gates, a Trotter decomposition is needed and the operator reads
\begin{equation}
\begin{split}
\hat{U}(\delta t)&=e^{-i\sum_{l} \hat{a}_{l}^{\dagger}\hat{a}_{l}\delta t}e^{i\sum_{l} (\hat{a}_l\hat{a}_{l+1}^{\dagger}+\hat{a}_{l}^{\dagger}\hat{a}_l)\delta t}e^{-i\sum_{l} \hat{a}_{l}^{\dagger}\hat{a}_{l}\delta t}\\
&=\hat{U}_{1}(\delta t)\hat{U}_{2}(\delta t)\hat{U}_{1}(\delta t)
\end{split}
\end{equation}
In order to demonstrate the process of operation more intuitively, we first introduce the explicit representation of state vectors and operators. We have
\begin{equation}
|0\rangle=\left(\begin{array}{c}
1  \\
0
\end{array}\right),
|1\rangle=\left(\begin{array}{c}
0  \\
1
\end{array}\right),
\end{equation}

\begin{equation}
\begin{split}
\hat{a}_{l}^{\dagger}&= |1\rangle_{Q_l}\langle0|\\
&=\left(\begin{array}{c}
0  \\
1
\end{array}\right)
\left(\begin{array}{cc}
1 & 0
\end{array}\right)\\
&=\left(\begin{array}{cc}
0 & 0\\
1 & 0
\end{array}\right),\\
\end{split}
\end{equation}

\begin{equation}
\begin{split}
\hat{a}_{l}&= |0\rangle_{Q_l}\langle1|\\
&=\left(\begin{array}{c}
1  \\
0
\end{array}\right)
\left(\begin{array}{cc}
0 & 1
\end{array}\right)\\
&=\left(\begin{array}{cc}
0 & 1\\
0 & 0
\end{array}\right),\\
\end{split}
\end{equation}
\begin{equation}
\begin{split}
\hat{a}_l\hat{a}^{\dag}_{l+1}&=|0\rangle_{Q_l}\langle1|\otimes|1\rangle_{Q_{l+1}}\langle0|\\
&=|01\rangle_{Q_{l,l+1}}\langle10|={\begin{pmatrix}0\\1\\0\\0\end{pmatrix}}(0 0 1 0)\\
&={\begin{pmatrix}0&0&0&0\\0&0&1&0\\0&0&0&0\\0&0&0&0\end{pmatrix}}\\
\end{split}
\end{equation}

\begin{equation}
\begin{split}
\hat{a}_l^{\dag}\hat{a}_{l+1}&=|1\rangle_{Q_l}\langle0|\otimes|0\rangle_{Q_{l+1}}\langle1|\\
&=|10\rangle_{Q_{l,l+1}}\langle01|\\
&={\begin{pmatrix}0\\0\\1\\0\end{pmatrix}}(0100)\\
&={\begin{pmatrix}0&0&0&0\\0&0&0&0\\0&1&0&0\\0&0&0&0\end{pmatrix}}
\end{split}
\end{equation}

Here, the time evolution operation is divided into two fundamental parts, a single-qubit(on-site term) operation $\hat{U}_{1}$ and a two-qubit(nn term) operation $\hat{U}_{2}$. Next, we are going to realize each operation by certain standard quantum gates. Starting from the single-qubit operation, $\hat{U}_1$ can be rewritten as
\begin{equation}
\begin{split}
\hat{U}_{1}(\delta t)&=e^{-i\sum_{l} \hat{a}_{l}^{\dagger}\hat{a}_{l}\delta t}\\
&=\prod_{l} e^{-i\hat{a}_{l}^{\dagger}\hat{a}_{l}\delta t}\\
&=\prod_{l} e^{-i \left(\begin{array}{cc}
	0 & 0 \\
	0 & 1
	\end{array}
	\right) \delta t}\\
&=\prod_{l}\left(\begin{array}{cc}
1 & 0 \\
0 & e^{-i\delta t}
\end{array}
\right)
\end{split}
\end{equation}
It is easy to find that $\hat{U}_{1}$ shares the similar configuration with the matrix representation of phase transition gate $P(\theta)$. Thus, by comparing the matrix representation of them, namely
\begin{equation}
\begin{split}
\hat{U}_{1}(\delta t)&=\prod_{l}\left(\begin{array}{cc}
1 & 0 \\
0 & e^{-i\delta t}
\end{array}
\right)\\
&=\prod_{l}\left(\begin{array}{cc}
1 & 0 \\
0 & e^{-i\theta}
\end{array}
\right)\\
&=\prod_{l} P(\theta)	,
\end{split}
\end{equation}
one can easily realize $\hat{U}_{1}$ operation with phase transition gate when the parameter of which $\theta=\delta t$.

In terms of the two-qubit operation, it is no longer so straightforward, but the idea of seizing equivalent effect after operating remains unchanged. We first illustrate the result of operating $\hat{U}_{2}$ on each two-qubit state. The matrix form of $\hat{U}_2$ reads

\begin{equation}
\begin{split}
\hat{U}_2(\delta t)=&e^{i\sum_{l}(\hat{a}_l\hat{a}_{l+1}^{\dag}+\hat{a}_l^{\dag}\hat{a}_{l+1})\delta t}\\=&\prod_le^{i(\hat{a}_l\hat{a}_{l+1}^{\dag}+\hat{a}_l^{\dag}\hat{a}_{l+1})\delta t}\\=&\prod_le^{i{\begin{pmatrix}0&0&0&0\\0&0&1&0\\0&1&0&0\\0&0&0&0\end{pmatrix}}_{l,l+1}\delta t}\\
=&\prod_l\left[\cos{\delta t}{\begin{pmatrix} 0&0&0&0\\0&1&0&0\\0&0&1&0\\0&0&0&0\\\end{pmatrix}}\right.\\
&\left.-i\sin{\delta t}{\begin{pmatrix} 0&0&0&0\\0&0&1&0\\0&1&0&0\\0&0&0&0\\\end{pmatrix}}+{\begin{pmatrix} 1&0&0&0\\0&0&0&0\\0&0&0&0\\0&0&0&1\\\end{pmatrix}}\right]_{l,l+1}.
\end{split}
\end{equation}
Applying the above representations inversely, we have
\begin{equation}\label{1}
\begin{split}
\hat{U}_2(\delta t)=&\cos{\delta t}(|01\rangle_{Q_{l,l+1}}\langle01|+|10\rangle_{Q_{l,l+1}}\langle10|)\\
&-i\sin{\delta t}(|01\rangle_{Q_{l,l+1}}\langle10|+|10\rangle_{Q_{l,l+1}}\langle01|)\\
&+|00\rangle_{Q_{l,l+1}}\langle00|+|11\rangle_{Q_{l,l+1}}\langle11|.\\
\end{split}
\end{equation}
Thus, it is pretty convenient to see the operation effect of $\hat{U}_2$ on different two-qubit eigenstates:
\begin{equation}\label{1}
\begin{split}
&\hat{U}_2(\delta t)|00\rangle=|00\rangle\\
&\hat{U}_2(\delta t)|01\rangle=\cos{\delta t}|01\rangle-i\sin{\delta t}|10\rangle\\
&\hat{U}_2(\delta t)|10\rangle=\cos{\delta t}|10\rangle-i\sin{\delta t}|01\rangle\\
&\hat{U}_2(\delta t)|11\rangle=|11\rangle\\
\end{split}
\end{equation}

Then, a series of equivalent two-qubit gates need to be found to realize the same effect. We use the combination of $Z^{\pm}$, $Y^{\pm}$ and $C-NOT$ gates to arrive at this goal, which is a general framework suitable for a variety of models. Note that, $Z^{\pm}$ and $Y^{\pm}$ gates can be realized by $R_z(\psi)$ and $R_y(\theta)$, and the relevant parameters can be obtained by comparing the following equations.
\begin{equation}
\begin{split}
Z^{+}&=e^{-i\frac{3}{4}\pi\sigma_z}\\
&=e^{-\frac{i}{2}\sigma_z\frac{3}{2}\pi}\\
&=R_z(\frac{3}{2}\pi) ,\\
\end{split}
\end{equation}

\begin{equation}
\begin{split}
Z^{-}&=e^{i\frac{3}{4}\pi\sigma_z}\\
&=e^{-\frac{i}{2}\sigma_z(-\frac{3}{2}\pi)}\\
&=R_z(-\frac{3}{2}\pi) ,\\
\end{split}
\end{equation}

\begin{equation}
\begin{split}
Y^{+}(\delta t)&=e^{-i\frac{\delta t}{2}\sigma_y}\\
&=R_y(\delta t), \\
\end{split}
\end{equation}

\begin{equation}
\begin{split}
Y^{-}(\delta t)&=e^{i\frac{\delta t}{2}\sigma_y}\\
&=e^{-i\frac{(-\delta t)}{2}\sigma_y}=R_y(-\delta t).\\
\end{split}
\end{equation}

Eventually, we successfully transform the time evolution operation under Heisenberg Hamiltonian into the language of quantum circuit.

\section*{B. Observing quantum walk on IBM cloud quantum computer}
After translating the time evolution into a series of quantum gates, we start to conduct digital experiments on IBM cloud quantum computers. Since IBM does not provide full access to the public and this is only a principle experiment, we build our dynamical circuit on a 5-qubit quantum computer.\\

%%%%%%
%%%%%%
\begin{figure*}[tbhp] \centering
	\includegraphics[width=0.9\textwidth]{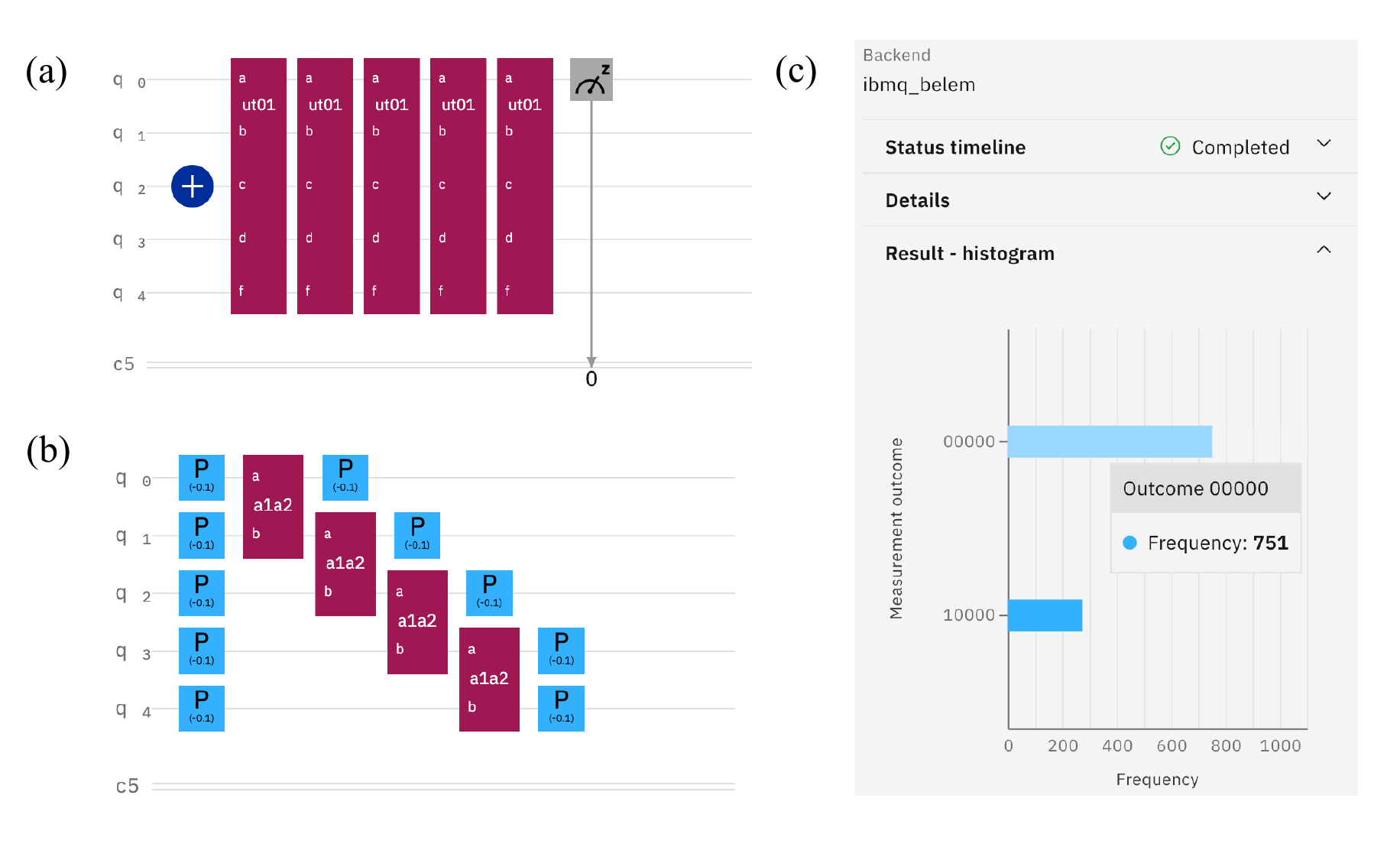}
	\caption{(Color online). The illustration of digital experiment on IBM cloud platform. (a) The circuit that simulates spin-flip dynamics in $0.5$ time period and conducts measurement on the first qubit. (b) The elements of the group 'ut01' in (a), which realizes $0.1$ time period of evolution. (c) The example of measurement result from the circuit.}\label{ibmoperation1}
\end{figure*}
%%%%%%
%%%%%%

To begin with, a NOT gate is put on the central qubit to conduct spin-flip. Then we place $R_z$ gates on each qubit to represent $\hat{U}_1$ operation and place a sequence of two-qubit gates on each two nn qubits to realize $\hat{U}_2$. For $\delta t=0.1$, the time evolution circuit of one $\delta t$ period is illustrated in Fig.~\ref{ibmoperation1}(b) ('$a1a2$' denotes the sequence of two-qubit gates in Fig.~\ref{ibmoperation1}(b)) and we take it as a group 'ut01'. Thus, we can demonstrate the evolution of single spin-flip by the step of $\delta t=0.1$ and measure the probablity density of each qubit (Fig.~\ref{ibmoperation1} (a) shows a $t=0.5$ evolution and a measurement on $q_0$).\\

Next, we can get the result of $1000$ measurements, which reveals the probablity density according to statistics rule [see Fig.~\ref{ibmoperation1}(c)]. By measuring the each qubit at various time points for many times and normalizing the data, we can obtain the density distribution at different key time point, which can be used to compare with our theoretical calculation.

\section*{C. Full data of the measured density distribution}
In this section, we demonstrate our experiment data of 5 qubits in 4 specific time points. The results are provided in TABLE.~\ref{tab:full data}.

\begin{table}[t]
\begin{tabular}{ccccccc}
\toprule[2pt]
 $\mathrm{t}=0$ & 1 & 2 & 3 & 4 & 5 & 6 \\
qubit$\backslash$ times & & & & & & \\
\midrule[1pt]$q_{0}$ & $0.00525$ & $0.00783$ & $0.00498$ & $0.00295$ & $0.00296$ & $0.00504$ \\
$q_{1}$ & $0.00840$ & $0.00685$ & $0.00498$ & $0.01083$ & $0.01281$ & $0.00504$ \\
$q_{2}$ & $0.97899$ & $0.97358$ & $0.97413$ & $0.96063$ & $0.96946$ & $0.97782$ \\
$q_{3}$ & $0.00420$ & $0.00685$ & $0.00398$ & $0.00492$ & $0.00591$ & $0.00101$ \\
$q_{4}$ & $0.00315$ & $0.00489$ & $0.01194$ & $0.02067$ & $0.00887$ & $0.01109$ \\
\bottomrule[2pt]\\
\toprule[2pt]
 $\mathrm{t}=0.25$ & 1 & 2 & 3 & 4 & 5 & 6 \\
qubit$\backslash$ times & & & & & & \\
\midrule[1pt]$q_{0}$ & $0.13477$ & $0.12764$ & $0.11037$ & $0.14517$ & $0.15214$ & $0.13477$ \\
$q_{1}$ & $0.18838$ & $0.21580$ & $0.23266$ & $0.22692$ & $0.22683$ & $0.18838$ \\
$q_{2}$ & $0.33805$ & $0.29690$ & $0.31022$ & $0.28823$ & $0.27939$ & $0.33805$ \\
$q_{3}$ & $0.19881$ & $0.21509$ & $0.19911$ & $0.19662$ & $0.19917$ & $0.19881$ \\
$q_{4}$ & $0.13999$ & $0.14457$ & $0.14765$ & $0.14306$ & $0.14246$ & $0.13999$ \\
\bottomrule[2pt]\\
\toprule[2pt]
$\mathrm{t}=0.50$ & 1 & 2 & 3 & 4 & 5 & 6 \\
qubit$\backslash$ times & & & & & & \\
\midrule[1pt]$q_{0}$ & $0.11119$ & $0.13988$ & $0.16210$ & $0.19623$ & $0.17516$ & $0.18953$ \\
$q_{1}$ & $0.26685$ & $0.27034$ & $0.27057$ & $0.23270$ & $0.21982$ & $0.22843$ \\
$q_{2}$ & $0.21682$ & $0.18023$ & $0.22243$ & $0.21887$ & $0.20447$ & $0.21287$ \\
$q_{3}$ & $0.23905$ & $0.23806$ & $0.23157$ & $0.20377$ & $0.24913$ & $0.25601$ \\
$q_{4}$ & $0.16609$ & $0.17149$ & $0.11335$ & $0.14843$ & $0.15143$ & $0.11315$ \\
\bottomrule[2pt]\\
\toprule[2pt]
 $\mathrm{t}=0.75$ & 1 & 2 & 3 & 4 & 5 & 6 \\
qubit$\backslash$ times & & & & & & \\
\midrule[1pt]$q_{0}$ & $0.22590$ & $0.21007$ & $0.20073$ & $0.19347$ & $0.19672$ & $0.21727$ \\
$q_{1}$ & $0.19639$ & $0.20147$ & $0.20133$ & $0.21096$ & $0.19915$ & $0.20800$ \\
$q_{2}$ & $0.17108$ & $0.17568$ & $0.19406$ & $0.19814$ & $0.17365$ & $0.18540$ \\
$q_{3}$ & $0.21265$ & $0.21744$ & $0.22074$ & $0.21620$ & $0.23497$ & $0.20510$ \\
$q_{4}$ & $0.19398$ & $0.19533$ & $0.18314$ & $0.18124$ & $0.19551$ & $0.18424$ \\
\bottomrule[2pt]\\

\end{tabular}
    \caption{Full data of density distribution from IBM quantum processor for $t=0,0.25,0.5,0.75$.}
    \label{tab:full data}
\end{table}
%%%%%%
%%%%%

\section*{S8. Visualized acceleration of quantum walk}
Please see the attached file of evolution.mp4


\begin{thebibliography}{73}
%1-3 many-body physics review

\bibitem{AZagoskin2014} A. Zagoskin, Quantum Theory of Many-Body Systems: Techniques and Applications, Springer, Berlin, 2014.

\bibitem{IBloch2008} I. Bloch, J. Dalibard, and W. Zwerger, Many-body physics with ultracold gases, Rev. Mod. Phys. \textbf{80}, 885 (2008).

\bibitem{LAmico2008} L. Amico, Rosario Fazio, Andreas Osterloh, and Vlatko Vedral, Entanglement in many-body systems, Rev. Mod. Phys. \textbf{80}, 517 (2008)




%-------------------------------------------------
%4-  quantum walks review
\bibitem{YAharonov1993} Y. Aharonov, L. Davidovich, and N. Zagury, Quantum random walks, Phys. Rev. A \textbf{48}, 1687 (1993).

\bibitem{NShenvi2003} N. Shenvi, J. Kempe, and K.B. Whaley, Quantum random-walk search algorithm, Phys. Rev. A \textbf0{67}, 052307 (2003).

\bibitem{AMChilds2009} A.M. Childs, Universal Computation by Quantum Walk, Phys. Rev. Lett.  \textbf{102}, 180501 (2009).

\bibitem{AMChilds2013} A.M. Childs, D. Gosset, Z. Webb, Universal Computation by Multiparticle Quantum Walk, Science \textbf{339}, 791 (2013).

\bibitem{SWZhang2003} S.-W. Zhang and H. Krakauer, Quantum Monte Carlo Method using Phase-Free Random Walks with Slater Determinants, Phys. Rev. Lett. \textbf{90}, 136401 (2003).

\bibitem{MSRudner2009} M.S. Rudner and L.S. Levitov, Topological Transition in a Non-Hermitian Quantum Walk, Phys. Rev. Lett. \textbf{106}, 065703 (2009).

\bibitem{SWZhang2011} S.-W. Zhang and H. Krakauer, Decoherence and Disorder in Quantum Walks: From Ballistic Spread to Localization, Phys. Rev. Lett. \textbf{106}, 180403 (2011).


\bibitem{LLehman2011} L. Lehman, V. Zatloukal, G.K. Brennen, J.K. Pachos, and Z.-H. Wang, Quantum Walks with Non-Abelian Anyons, Phys. Rev. Lett. \textbf{122}, 230404(2011).

\bibitem{IVakulchyk2019} I. Vakulchyk, M.V. Fistul, and S. Flach, Wave Packet Spreading with Disordered Nonlinear Discrete-Time Quantum Walks, Phys. Rev. Lett. \textbf{122}, 040501 (2019).


\bibitem{HChalabi2019} H. Chalabi, S. Barik, S. Mittal, T.E. Murphy, M. Hafezi, and E. Waks, Synthetic Gauge Field for Two-Dimensional Time-Multiplexed Quantum Random Walks, Phys. Rev. Lett. \textbf{123}, 150503 (2019)

\bibitem{SChakraborty2020} S. Chakraborty, K. Luh, and J. Roland, How Fast Do Quantum Walks Mix?, Phys. Rev. Lett. \textbf{124}, 050501 (2020) .

\bibitem{LKUpreti2020} L.K. Upreti, C. Evain, S. Randoux, P. Suret, A. Amo, and P. Delplace, Topological Swing of Bloch Oscillations in Quantum Walks, Phys. Rev. Lett. \textbf{125}, 186804 (2020)

\bibitem{PWrzosek2020} P. Wrzosek, K. Wohlfeld, D. Hofmann, T. Sowi\'{n}ski, and M. A. Sentef, Quantum walk versus classical wave: Distinguishing ground states of quantum magnets by spacetime dynamics, Phys. Rev. B \textbf{102}, 024440(2020).

\bibitem{KManouchehri2014}  K. Manouchehri, J. Wang, Physical Implementation of Quantum Walks (Springer-Verlag, Berlin, 2014).


%-----------------------------------
%photonic system

\bibitem{HBPerets2008} H.B. Perets, Y. Lahini, F. Pozzi, M. Sorel, R. Morandotti, and Y. Silberberg, Realization of Quantum Walks with Negligible Decoherence in Waveguide Lattices, Phys. Rev. Lett. \textbf{100}, 170506 (2008).

\bibitem{ASchreiber2010} A. Schreiber, K.N. Cassemiro, V. Poto\v{c}ek, A.G\'{a}bris, P. J. Mosley, E. Andersson, I. Jex, and Ch. Silberhorn, Photons Walking the Line: A Quantum Walk with Adjustable Coin Operations, Phys. Rev. Lett. \textbf{104}, 050502 (2010).

\bibitem{MABroome2010} M.A. Broome, A. Fedrizzi, B.P. Lanyon, I. Kassal, A. Aspuru-Guzik, and A.G. White, Discrete Single-Photon Quantum Walks with Tunable Decoherence, Phys. Rev. Lett. \textbf{104}, 153602 (2010).

\bibitem{LSansoni2012} L. Sansoni, F. Sciarrino, G. Vallone, P. Mataloni, A. Crespi, R. Ramponi, and R. Osellame, Two-Particle Bosonic-Fermionic Quantum Walk via Integrated Photonics, Phys. Rev. Lett. \textbf{108}, 010502 (2012).

\bibitem{SChakraborty2016} S. Chakraborty, L. Novo, A. Ambainis, and Y. Omar, Spatial Search by Quantum Walk is Optimal for Almost all Graphs, Phys. Rev. Lett. \textbf{116}, 100501 (2016).

\bibitem{KKWang2019} K.-K. Wang, X.Z. Qiu, L. Xiao, X. Zhan, Z. H. Bian, W. Yi, and P. Xue, Simulating Dynamic Quantum Phase Transitions in Photonic Quantum Walks, Phys. Rev. Lett. \textbf{122}, 020501 (2019).

\bibitem{TWu2020} T. Wu, J.A. Izaac, Z.-X. Li, K. Wang, Z.-Z. Chen, S.-N. Zhu, J.B. Wang, and X.-S. Ma, Experimental Parity-Time Symmetric Quantum Walks for Centrality Ranking on Directed Graphs, Phys. Rev. Lett. \textbf{125}, 240501 (2020).

\bibitem{YWang2020} Y. Wang, Z.-W. Cui, Y.-H. Lu, X.-M. Zhang, J. Gao, Y.-J. Chang, M.-H. Yung, and X.-M. Jin, Integrated Quantum-Walk Structure and NAND Tree on a Photonic Chip, Phys. Rev. Lett. \textbf{125}, 160502 (2020)

\bibitem{LXiao2021p1} L. Xiao, T.-S. Deng, K.-K. Wang, Z. Wang, W. Yi, and P. Xue, Observation of Non-Bloch Parity-Time Symmetry and Exceptional Points, Phys. Rev. Lett. \textbf{126}, 230402 (2021).

\bibitem{LXiao2021p2} L. Xiao, D.-K. Qu, K.-K. Wang, H.-W. Li, J.-Y. Dai, B. D\'{o}ra, M. Heyl, R. Moessner, W. Yi, and P. Xue, Non-Hermitian Kibble-Zurek Mechanism with Tunable Complexity in Single-Photon Interferometry, PRX Quantum \textbf{2}, 020313 (2021).

\bibitem{KKWang2021} K.-K. Wang, L. Xiao, J.C. Budich, W. Yi, and P. Xue, Simulating Exceptional Non-Hermitian Metals with Single-Photon Interferometry, Phys. Rev. Lett. \textbf{127}, 026404 (2021)

\bibitem{AGeraldi2019} A. Geraldi, A. Laneve, L.D. Bonavena, et al.,  Experimental Investigation of Superdiffusion via Coherent Disordered Quantum Walks, Phys. Rev. Lett. \textbf{123}, 140501 (2019).


\bibitem{TGiordani2019} T. Giordani, E. Polino, S. Emiliani, et al., Experimental Engineering of Arbitrary Qudit States with Discrete-Time Quantum Walks, Phys. Rev. Lett. \textbf{122}, 020503 (2019).

\bibitem{SBarkhofen2018}
S. Barkhofen, L. Lorz, T. Nitsche, C. Silberhorn, and H. Schomerus,  Supersymmetric Polarization Anomaly in Photonic Discrete-Time Quantum Walks, Phys. Rev. Lett. \textbf{121}, 260501 (2018).

\bibitem{BWang2018} B. Wang, T. Chen, and X.-D. Zhang, Experimental Observation of Topologically Protected Bound States with Vanishing Chern Numbers in a Two-Dimensional Quantum Walk, Phys. Rev. Lett. \textbf{121}, 100501 (2018)

\bibitem{CChen2018} C. Chen, X. Ding, J. Qin, Y. He, et al., Observation of Topologically Protected Edge States in a Photonic Two-Dimensional Quantum Walk, Phys. Rev. Lett. \textbf{121}, 100502 (2018).



\bibitem{XYXu2018} X.-Y. Xu, Q.-Q. Wang, W.-W. Pan, et al., Measuring the Winding Number in a Large-Scale Chiral Quantum Walk, Phys. Rev. Lett. \textbf{120}, 260501 (2018).

\bibitem{XGQiang2021} X.-G. Qiang, Y.-Z Wang, S.-C. Xue, Implementing graph-theoretic quantum algorithms on a silicon photonic quantum walk processor, Sci. Adv. \textbf{7}, eabb8375 (2021).

\bibitem{HTang2018} H. Tang, X.-F. Lin, Z. Feng, J.-Y. Chen, et al., Experimental two-dimensional quantum walk on a photonic chip, Sci. Adv. \textbf{4}, eaat3174 (2018).


\bibitem{XGQiang2018} X.-G. Qiang, X.-Q. Zhou, J.-W. Wang, et al., large-scale silicon quantum photonics implementing arbitrary two-qubit processing, Nat. Photon. \textbf{12}, 534 (2018).


\bibitem{LXiao2017} L. Xiao, X. Zhan, Z.H. Bian, et al., Observation of topological edge states in parity-time-symmetric quantum walks, Nat. Phys. \textbf{13}, 1117 (2017).

\bibitem{XYXu2021} X.-Y. Xu, X.-W. Wang, D.-Y. Chen, C.M. Smith, and X.-M. Jin, Quantum transport in fractal networks,  Nat. Photon. (2021). https://doi.org/10.1038/s41566-021-00845-4


%-----------------------------------------------
%trapped ion
%PXue2009,RMatjeschk2012,HSchmitz2009,Zahringer2010,MTamura2020

REPORT
\bibitem{PXue2009} P. Xue, B. C. Sanders, and D. Leibfried, Quantum Walk on a Line for a Trapped Ion, Phys. Rev. Lett. \textbf{103}, 183602 (2009).

\bibitem{RMatjeschk2012} R. Matjeschk, Ch. Schneider, M. Enderlein, et al., Experimental simulation and limitations of quantum walks with trapped ions, New J. Phys. \textbf{14}, 035012 (2012)



\bibitem{HSchmitz2009} H. Schmitz, R. Matjeschk, Ch. Schneider, J. Glueckert, M. Enderlein, T. Huber, and T. Schaetz, Quantum Walk of a Trapped Ion in Phase Space, Phys. Rev. Lett. \textbf{103}, 090504 (2009).

\bibitem{Zahringer2010} F. Z\"{a}hringer, G. Kirchmair, R. Gerritsma, E. Solano, R. Blatt, and C. F. Roos, Realization of a Quantum Walk with One and Two Trapped Ions, Phys. Rev. Lett. \textbf{104}, 100503 (2010).

\bibitem{MTamura2020} M. Tamura, T. Mukaiyama, and K. Toyoda, Quantum Walks of a Phonon in Trapped Ions, Phys. Rev. Lett. \textbf{124}, 200501 (2020).

%------------------------------------------------------------------
%Cold atoms
%MKarski2009,CWeitenberg2011,Tfukuhara2013,DXXie2020,JCarlstrom2016;VVRamasesh2017,ZGYan2019,MGong2021

\bibitem{MKarski2009} M. Karski, L. F\"{o}rster, J.-M. Choi, et al., Quantum Walk in Position Space with Single Optically Trapped Atoms, Science \textbf{325}, 174 (2009).

\bibitem{CWeitenberg2011} C. Weitenberg, M. Endres, J.F. Sherson, M. Cheneau, P. Schau? T. Fukuhara, I. Bloch, S. Kuhr, Single-spin addressing in an atomic Mott insulator. Nature \textbf{471}, 319 (2011).

\bibitem{TFukuhara2013} T. Fukuhara, P. Schau? M. Endres, S. Hild, M. Cheneau, I. Bloch, C. Gross, Microscopic observation of magnon bound states and their dynamics. Nature \textbf{502}, 76 (2013).

\bibitem{DXXie2020} D.-X. Xie, T.-S. Deng, T. Xiao, W. Gou, T. Chen, W. Yi, and B. Yan, Topological Quantum Walks in Momentum Space with a Bose-Einstein Condensate, Phys. Rev. Lett. \textbf{124}, 050502 (2020).

\bibitem{JCarlstrom2016} J. Carlstr\"{o}m, N. Prokof'ev, and B. Svistunov, Quantum Walk in Degenerate Spin Environments, Phys. Rev. Lett. \textbf{116}, 247202 (2016).


%--------------------------------------------------------
%Superconducting circuits
\bibitem{VVRamasesh2017} V.V. Ramasesh, E. Flurin, M. Rudner, I. Siddiqi, and N.Y. Yao, Direct Probe of Topological Invariants Using Bloch Oscillating Quantum Walks, Phys. Rev. Lett. \textbf{118}, 130501 (2017).

\bibitem{ZGYan2019} Z.-G. Yan, Y.-R. Zhang, M. Gong, et al.,  Strongly correlated quantum walks with a 12-qubit superconducting processor, Science \textbf{364}, 753 (2019).

\bibitem{MGong2021} M. Gong, S.-Y Wang, C. Zha, et al., Quantum walks on a programmable two-dimensional 62-qubit superconducting processor, Science \textbf{372}, 948 (2021).

\bibitem{YWu2021} Y. Wu, W.-S. Bao, S. Cao, et al., Strong Quantum Computational Advantage Using a Superconducting Quantum Processor, Phys. Rev. Lett. \textbf{127}, 180501 (2021).


%=======================================
%



%??
%many-body dynamics

%MCheneau2012,JZhang2017,MCTran2020,CMonroe2021,PIlzhofer2021
\bibitem{MCheneau2012} M. Cheneau, P. Barmettler, D. Poletti, et  al., Light-cone-like spreading of correlations in a quantum many-body system, Nature \textbf{481},  484 (2012).

\bibitem{JZhang2017} J. Zhang, G. Pagano, P.W. Hess, A. Kyprianidis, P. Becker, H. Kaplan, A.V. Gorshkov, Z.-X. Gong and C. Monroe, Observation of a many-body dynamical phase transition with a 53-qubit quantum simulator, Nature \textbf{551}, 601 (2017).

\bibitem{MCTran2020} M.C. Tran, C.-F. Chen, A. Ehrenberg, Hierarchy of Linear Light Cones with Long-Range Interactions, Phys. Rev. X \textbf{10}, 031009 (2020).

\bibitem{CMonroe2021} C. Monroe, W.C. Campbell, L.-M. Duan, et al, Programmable quantum simulations of spin systems with trapped ions, Rev. Mod. Phys. \textbf{93}, 025001 (2021).



\bibitem{PIlzhofer2021} P. Ilzh\"{o}fer, M. Sohmen, G. Durastante, et  al., Phase coherence in out-of-equilibrium supersolid states of ultracold dipolar atoms, Nat. Phys. \textbf{17}, 356 (2021).




%-----------------
%otoc
%-------------------------
%OTOCs, scrambling and many-body dynamics
%\cite{JLi2017,MGarttner2017,ANahum2018,VKhemani2018,SVSyzranov2018,MHeyl2018,MGarttner2018,BYoshida2019,CMurthy2019,KALandsman2019,SLXu2020,SChoi2020,RJLewis-Swan2020}


\bibitem{JLi2017} J. Li, R.-H. Fan, H.-Y. Wang, B.-T. Ye, B. Zeng, H. Zhai, X.-H. Peng, and J.-F. Du, Measuring Out-of-Time-Order Correlators on a Nuclear Magnetic Resonance Quantum Simulator, Phys. Rev. X \textbf{7}, 031011 (2017).

\bibitem{MGarttner2017} M. G\"{a}rttner, J.G. Bohnet, A. Safavi-Naini, M.L. Wall, J.J. Bollinger and A.M. Rey, Measuring out-of-time-order correlations and multiple quantum spectra in a trapped-ion quantum magnet, Nat. Phys. \textbf{13}, 781 (2017).

\bibitem{ANahum2018} A. Nahum, S. Vijay, and J. Haah, Operator Spreading in Random Unitary Circuits, Phys. Rev. X \textbf{8}, 021014 (2018).

\bibitem{VKhemani2018} V. Khemani, A. Vishwanath, and D.A. Huse, Operator Spreading and the Emergence of Dissipative Hydrodynamics under Unitary Evolution with Conservation Laws, Phys. Rev. X \textbf{8}, 031057 (2018).

\bibitem{SVSyzranov2018} S.V. Syzranov, A.V. Gorshkov, and V. Galitski, Out-of-time-order correlators in finite open systems, Phys. Rev. B \textbf{97}, 161114(R) (2018).

\bibitem{MHeyl2018} M. Heyl, F. Pollmann, and B. D\'{o}ra, Detecting Equilibrium and Dynamical Quantum Phase Transitions in Ising Chains via Out-of-Time-Ordered Correlators, Phys. Rev. Lett. \textbf{121}, 016801 (2018).

\bibitem{MGarttner2018} M. G\"{a}rttner, P. Hauke, and A.M. Rey, Relating Out-of-Time-Order Correlations to Entanglement via Multiple-Quantum Coherences, Phys. Rev. Lett. \textbf{120}, 040402 (2018).

\bibitem{BYoshida2019} B. Yoshida and N.Y. Yao, Disentangling Scrambling and Decoherence via Quantum Teleportation, Phys. Rev. X \textbf{9}, 011006 (2019).

\bibitem{CMurthy2019} C. Murthy and M. Srednicki, Bounds on Chaos from the Eigenstate Thermalization Hypothesis, Phys. Rev. Lett. \textbf{123}, 230606 (2019).

\bibitem{KALandsman2019} K.A. Landsman, C. Figgatt, T. Schuster, N.M. Linke, B. Yoshida, N.Y. Yao and C. Monroe, Verified quantum information scrambling, Nature \textbf{567}, 61 (2019).

\bibitem{SLXu2020} S.L. Xu and B. Swingle, Accessing scrambling using matrix product operators, Nat. Phys. \textbf{16}, 199 (2020).

\bibitem{SChoi2020} S. Choi, Y. Bao, X.-L. Qi, and E. Altman,  Quantum error correction in scrambling dynamics and measurement-induced phase transition, Phys. Rev. Lett. \textbf{125}, 030505 (2020).

\bibitem{RJLewis-Swan2020} R.J. Lewis-Swan, S.R. Muleady, and A.M. Rey, Detecting Out-of-Time-Order Correlations via Quasiadiabatic Echoes as a Tool to Reveal Quantum Coherence in Equilibrium Quantum Phase Transitions, Phys. Rev. Lett. \textbf{125}, 240605 (2020).








%localized system
%~\cite{KXWei2018,SLXu2019,MMcGinley2019,PRoushan2017,HTShen2017,KSlagle2017,MCTran2019,KXu2018,NYYao2014}


\bibitem{KXWei2018} K.X. Wei, C. Ramanathan, and P. Cappellaro, Exploring Localization in Nuclear Spin Chains, Phys. Rev. Lett. \textbf{120}, 070501(2018).

\bibitem{SLXu2019} S.L. Xu and B. Swingle, Locality, Quantum Fluctuations, and Scrambling, Phys. Rev. X \textbf{9}, 031048 (2019).

\bibitem{MMcGinley2019} M. McGinley, A. Nunnenkamp, and J. Knolle, Slow Growth of Out-of-Time-Order Correlators and Entanglement Entropy in Integrable Disordered Systems, Phys. Rev. Lett. \textbf{122}, 020603 (2019).

\bibitem{PRoushan2017} P. Roushan, C. Neill, J. Tangpanitanon, et al., Spectroscopic signatures of localization
with interacting photons in superconducting qubits, Science \textbf{358}, 1175 (2017).

\bibitem{HTShen2017} H.T. Shen, P.F. Zhang, R.H. Fan, and H. Zhai, Out-of-time-order correlation at a quantum phase transition, Phys. Rev. B \textbf{96}, 054503 (2017).

\bibitem{YHuang2016} Y. Huang, Y.-L. Zhang, and X. Chen, Out-of-time-ordered correlators in many-body localized systems, Ann. Phys. (Amsterdam) \textbf{529}, 1600318 (2016).,
\bibitem{KSlagle2017} K. Slagle, Z. Bi, Y.-Z. You, and C.K. Xu, Out-of-time-order correlation in marginal many-body localized systems, Phys. Rev. B \textbf{95}, 165136 (2017).
\bibitem{MCTran2019} M.C. Tran, A.Y. Guo, Y. Su, J.R. Garrison, Z. Eldredge, M. Foss-Feig, A.M. Childs, and A.V. Gorshkov, Locality and Digital Quantum Simulation of Power-Law Interactions, Phys. Rev. X \textbf{9}, 031006 (2019).

\bibitem{KXu2018} K. Xu, J.-J. Chen, Y. Zeng, et al., Emulating Many-Body Localization with a Superconducting Quantum Processor, Phys. Rev. Lett. \textbf{120}, 050507 (2018).

\bibitem{NYYao2014} N.Y. Yao, C.R. Laumann, S. Gopalakrishnan, M. Knap, M. Mueller, E.A. Demler, and M.D. Lukin, Many-Body Localization in Dipolar Systems, Phys. Rev. Lett. \textbf{113}, 243002 (2014).


%Chaotic physics ~\cite{EBRozenbaum2017,IGarcia-Mata2018,BBertini2019,BYan2020,DEParker2019,JRammensee2018,TRXu2020,WLZhao2021}

\bibitem{EBRozenbaum2017} E.B. Rozenbaum, S. Ganeshan, and V. Galitski, Lyapunov Exponent and Out-of-Time-Ordered Correlator's Growth Rate in a Chaotic System, Phys. Rev. Lett. \textbf{118}, 086801 (2017).

\bibitem{IGarcia-Mata2018} I. Garc\'{i}a-Mata, M. Saraceno, R.A. Jalabert, A.J. Roncaglia, and D.A. Wisniacki, Chaos Signatures in the Short and Long Time Behavior of the Out-of-Time Ordered Correlator, Phys. Rev. Lett. \textbf{121}, 210601 (2018).

\bibitem{BBertini2019} B. Bertini, P. Kos, and T. Prosen, Entanglement Spreading in a Minimal Model of Maximal Many-Body Quantum Chaos, Phys. Rev. X \textbf{9}, 021033 (2019).

\bibitem{BYan2020} B. Yan, L. Cincio, and W.H. Zurek, Information Scrambling and Loschmidt Echo, Phys. Rev. Lett. \textbf{124}, 160603 (2020).

\bibitem{DEParker2019} D.E. Parker, X.Y. Cao, A. Avdoshkin, T. Scaffidi, and E. Altman, A Universal Operator Growth Hypothesis, Phys. Rev. X \textbf{9}, 041017 (2019).

\bibitem{JRammensee2018} J. Rammensee, J.D. Urbina, and K. Richter, Many-Body Quantum Interference and the Saturation of Out-of-Time-Order Correlators, Phys. Rev. Lett. \textbf{121}, 124101 (2018).

\bibitem{TRXu2020} T.R. Xu, T. Scaffidi, and X.Y. Cao, Does Scrambling Equal Chaos?, Phys. Rev. Lett. \textbf{124}, 140602 (2020).

\bibitem{WLZhao2021} W.-L. Zhao, Y. Hu, Z. Li, and Q. Wang, Super-exponential growth of out-of-time-ordered correlators, Phys. Rev. B \textbf{103}, 184311 (2021).


%Black Hole physics ~\cite{DARoberts2015,JMaldacena2016,JdeBoer2018}
\bibitem{DARoberts2015} D.A. Roberts and D. Stanford, Diagnosing Chaos Using Four-Point Functions in Two-Dimensional Conformal Field Theory, Phys. Rev. Lett. \textbf{115}, 131603 (2015).

\bibitem{JMaldacena2016} J. Maldacena, S. H. Shenker, and D. Stanford, A bound on chaos, J. High Energy Phys. \textbf{08}, 106 (2016).

\bibitem{JMaldacena1999} J. Maldacena, The Large-N Limit of Superconformal Field Theories and Supergravity, Int. J. of Theor. Phys., \textbf{38}, 1113 (1999).


\bibitem{JdeBoer2018} J. de Boer, E. Llabr\'{e}s, J.F. Pedraza, and D. Vegh, Chaotic Strings in AdS/CFT, Phys. Rev. Lett. \textbf{120}, 201604 (2018).

%topological system

\bibitem{CBDag2019a} C.B. Da\v{g}, K. Sun, and L.-M. Duan, Detection of Quantum Phases via Out-of-Time-Order Correlators, Phys. Rev. Lett. \textbf{123}, 140602 (2019).


\bibitem{CBDag2019b} C.B. Da\v{g} and L.-M. Duan, Detection of out-of-time-order correlators and information scrambling in cold atoms: Ladder-XX model, Phys. Rev. A \textbf{99}, 052322 (2019).



\bibitem{suppM} Supplementary Materials I: Detailed linear spin wave method, magnon dynamics, otoc and butterfly velocities are provided in S1-S3. S4 provide the comparison of OTOCs between theoretical predictions and the results of numerical Exact diagonalization.

Supplementary Materials II: Movie of visualized quantum walks. https://youtu.be/xX1M-N0CM9o




%Lieb-Robinson bounds
%EHLieb1972,SBravyi2006,BNachtergaele2006,MCheneau2012,NLashkari2013,JJunemann2013,MFoss-Feig2015,DMStorch2015,DMStorch2015,TMatsuta2017,DVElse2020,TKuwahara2020}
\bibitem{EHLieb1972} E.H. Lieb and D.W. Robinson, The finite group velocity of quantum spin systems, Commun. Math. Phys. \textbf{28}, 251 (1972).

\bibitem{SBravyi2006} S. Bravyi, M.B. Hastings and F. Verstraete, Lieb-Robinson bounds and the generation of correlations and topological quantum order, Phys. Rev. Lett. \textbf{97}, 050401 (2006).

\bibitem{BNachtergaele2006} B. Nachtergaele, R. Sims, Lieb-Robinson bound and the exponential clustering theorem, Commun. Math. Phys. \textbf{265}, 119 (2006).



\bibitem{NLashkari2013} N. Lashkari, D. Stanford, M. Hastings, T. Osborne, and P. Hayden, Towards the fast scrambling conjecture, J. High Energy Phys. \textbf{04}, 022 (2013).

\bibitem{JJunemann2013} J. J\"{u}nemann, A. Cadarso, D. Perez-Garcia, A. Bermudez and J.J. Garcia-Ripoll, Lieb-Robinson bounds for spin-boson lattice models and trapped ions, Phys. Rev. Lett. \textbf{111}, 230404 (2013)

\bibitem{MFoss-Feig2015} M. Foss-Feig, Z.-X. Gong, C.W. Clark, and A. V. Gorshkov, Nearly Linear Light Cones in Long-Range Interacting Quantum Systems, Phys. Rev. Lett. \textbf{114}, 157201 (2015).

\bibitem{DMStorch2015} D.-M. Storch, , M.V.D. Worm, and M. Kastner, Interplay of soundcone and supersonic propagation in lattice models with power law interactions, New J. Phys. \textbf{17}, 063021 (2015).

\bibitem{CCChien2015} C.-C. Chien, S. Peotta and M.D. Ventra, Quantum transport in ultracold atoms, Nat. Phys. \textbf{11}, 998 (2015).

\bibitem{TMatsuta2017} T. Matsuta, T. Koma, and S. Nakamura, Improving the Lieb-Robinson bound for long-range interactions, Ann. Henri Poincar\'{e} \textbf{18}, 519 (2017).

\bibitem{DVElse2020} D.V. Else, , F. Machado, C. Nayak, and N.Y. Yao, Improved Lieb-Robinson bound for many-body Hamiltonians with power-law interactions, Phys. Rev. A \textbf{101}, 022333 (2020).

\bibitem{TKuwahara2020} T. Kuwahara and K. Saito, Strictly Linear Light Cones in Long-Range Interacting Systems of Arbitrary Dimensions, Phys. Rev. X \textbf{10}, 031010 (2020).

\bibitem{XMi2021}X. Mi, P. Roushan, C. Quintana, Information scrambling in quantum circuits, Science 10.1126/science.abg5029(2021).

\bibitem{IBM127news}https://siliconangle.com/2021/11/15/ibm-debuts-new-eagle-quantum-processor-127-qubits/

\end{thebibliography}

\begin{thebibliography}{73}
	
\bibitem{DVElse2020} D.V. Else, , F. Machado, C. Nayak, and N.Y. Yao, Improved Lieb-Robinson bound for many-body Hamiltonians with power-law interactions, Phys. Rev. A \textbf{101}, 022333 (2020).

\bibitem{TKuwahara2020} T. Kuwahara and K. Saito, Strictly Linear Light Cones in Long-Range Interacting Systems of Arbitrary Dimensions, Phys. Rev. X \textbf{10}, 031010 (2020).





\end{thebibliography}
\end{document}